\documentclass[preprint,3p,10pt,times,onecolumn]{elsarticle}
\usepackage{lineno}
\usepackage{latexsym}
\AtEndDocument{\clearpage}

\usepackage{dcolumn}
\usepackage{amssymb}
\usepackage{amsmath}
\usepackage{amsfonts}
\usepackage{mathtools}
\usepackage{graphicx}
\usepackage{multirow}
\usepackage{subfigure}
\usepackage{mathtools}
\usepackage{rotating}
\usepackage{adjustbox}
\usepackage{rotating}
\usepackage{adjustbox}
\usepackage{blindtext}
\usepackage{booktabs}

\begin{document}
\begin{frontmatter}


\title{
Asymmetric return rates and wealth distribution influenced by the introduction of technical analysis into a behavioral agent based model
}

%

\author{F.M. ~Stefan\corref{cor1}\fnref{fn1}}
\ead{fischer@dppg.cefetmg.br}
\author{A.P.F. ~Atman\corref{cor2}\fnref{fn2}}
\ead{atman@cefetmg.br}

\cortext[cor1]{Principal corresponding author}
\cortext[cor2]{Corresponding author}

\address[fn1]{Federal Center for Technological Education of Minas Gerais - CEFET--MG, Av. Amazonas 7675, 30510-000, Belo Horizonte-MG, Brazil.}
\address[fn2]{Departamento de F\'{\i}sica e Matem\'atica and National Institute of Science and Technology for Complex Systems,
Federal Center for Technological Education of Minas Gerais - CEFET--MG. Av. Amazonas 7675, 30510-000, Belo Horizonte-MG, Brazil.}

%
%


\begin{abstract}
Behavioral Finance has become a challenge to the scientific community. Based on the assumption that behavioral aspects of investors may explain some features of the Stock Market, we propose an agent based model to study quantitatively this relationship. In order to approximate the simulated market to the complexity of real markets, we consider that the investors are connected among them through a small world network; each one has its own psychological profile (Imitation, Anti-Imitation, Random); two different strategies for decision making: one of them is based on the trust neighborhood of the investor and the other one  considers a technical analysis, the momentum of the market index technique. We analyze the market index fluctuations, the wealth distribution of the investors according to their psychological profiles and the rate of return distribution. Moreover, we analyze the influence of changing the psychological profile of the hub of the network and report interesting results which show how and when anti-imitation becomes the most profitable strategy for investment. Besides this, an intriguing asymmetry of the return rate distribution is explained considering the behavioral aspect of the investors. This asymmetry is quite robust being observed even when a completely different algorithm to calculate the decision making of the investors was applied to it, a remarkable result which, up to our knowledge, has never been reported before.

\end{abstract}



\end{frontmatter}


\section{Introduction}

In the last decades many researchers have devoted their studies to understand the financial market expecting that it behaves as a complex system. Different branches of the science  started researching how the behavior of the investors work in real systems as an attempting to understand it as a whole \cite{Boccara,Bouchaud,lebaron06,lebaron94,lebaron99,cajueiro04,tanyaraujo,fudenberg,Hart,Coolen,Lux2005,Lux2009,Lux2012}. In particular, we have the Economic system which works as a complex system, where people, companies and markets are at the microscopic level trying to increase their profits predicting the behavior of the investors \cite{Mitchell,sornette,mantegna,kirman10}. We already know that many situations affect the stock market such as the value of the index which depends on the choice of the investors to either buy, sell or hold their stocks. If this decision of the investors is based only on the behavior of their trust network, psychological tendencies arise from various processes \cite{Amos}. Then, the trust neighborhood and also the market index behavior are driving forces in order to influence the decision making of the investors. 

Financial Market (FM) is the place where the financial assets are traded \footnote{although, the assets do not, necessarily, have to be traded in a market}. There are some economic functions provided by the FM which show how the finance is related to the economy system. The three major functions are: the interactions among buyers and sellers which determine the price of the traded asset, which means that they determine the required return on a financial asset; liquidity which is a mechanism that FM provides to investors to sell their financial assets and it can distinguish different kinds of markets (liquidity degree); and the last one is related to reducing the search costs of transactions (the money spent to advertise the one's intention to sell or purchase a financial asset) and information costs (assessing the investment merits of a financial asset: the amount and the likelihood of the cash flow to be generated). The participants of the FM can be households, business entities, national government agencies, supranational (World Bank, European Investment Bank) \textit{etc} \cite{Fabozzi}. 

Stock market has a large number of interacting agents making decisions all the time \cite{lebaron06,Lux1,Lux2005,Lux2009}. Many researchers are trying to study its behavior based on the similarity with a complex system. It has no central controller and the dynamic observed from it brings some patterns which are hard to predict such as patterns of bubbles and crashes. These patterns are good examples of the capacity of the market to exhibit a self-organization behavior and then emergent properties \cite{Bouchaud,sornette,Mitchell}. 

Duncan J. Watts e Steven Strogatz show that the average distance $\langle l\rangle$ between two nodes increases increases logarithmically with the size of the network. We see that $\langle l\rangle=\ln N$ \cite{Strogatz,Newman}. This characteristic can also be seen in free scale network (SFN). Then, we can infer that most   SFN are a kind of small world network (SWN) \cite{cohen2010,Newman}. Therefore, we can represent our society of investors in the stock market connected through a Scale Free Network as shown in the Figure \ref{fig:swn}.

Several researchers from social  to computation science have shown that the social networks are scale free. This complex network has been used as a tool to connect people in the real world \cite{Philip,santos,Newman}. As a consequence, it is able to describe social interactions and it is very important to the spread of information, playing a central role on the social relations \cite{Newman}. In this way, we use this scale free network (SFN) to represent our society of investors in a financial market which makes the most realistic  connection among the investors, building their trust neighborhood \cite{Philip,santos,Barabasi,Tiziana1}. One of the most important  characteristics of the SWN is that there is almost always a shortcut connecting any two nodes. In this situation, there are many alternative routes between any two points, and it is very likely that some will involve only a few jumps (links). Moreover, a SFN feature is that highly connected nodes have a greater-than-average chance of being linked to other highly connected nodes creating hubs which are highly connected among themselvs. Members of a hub trust network might, for instance, share information, quickly synchronizing their ``cluster'',  while for the rest of network information percolates slowly  by local interactions \cite{Philip,Bonan,Newman_2000,Newman}.

An agent based model grounded on behavioral stochastic Cellular Automata (CA) has been implemented in order to reproduce the main features of the Stock Market and  study it as a complex system \cite{Stefan}. However, to approach the agents to real investors, it was mandatory to consider a technical analysis in the decision making of the agents, feature which has not been considered in the algorithm  yet. As forecasting price movements is the core of the technical analysis, this methodology uses past prices,volume and/or open interest in order to bring several kinds of forecasting techniques such as chart analysis taking into account shapes in bar charts, as gaps, spikes, flags, etc, which tests the profitability of visual chart patterns, cycle analysis and computerized technical trading systems \cite{Steven,Pring,Murphy,Edwards,Irwin,Lux2005}. Technical analysis is widely used among traders and financial professionals (i.e. the participants of the FM), and is very often used by active day traders and market makers \cite{Irwin}.
 
Thus, in this work we improve the original model \cite{Stefan} exploring the decision making algorithm which is now combining two strategies to help the investor to make an investment. At each time step, the investor will consider his trust neighborhood and the trend of the stock market index with different weights. In order to study the stock market index tendency, we are going to apply a technical analysis methodology called momentum. This indicator warns about latent strengths or weaknesses in the tendency by monitoring the price \cite{Steven,Pring,Murphy}. Yet, it will be taken into account the psychological behavior of each investor when applying these two strategies. In this way, the investor will take a decision of either buying, holding or selling stocks based on combination of his trust neighborhood and momentum technique choices.

Next section, we present the methodology developed to implement the rules through the CA, the Complex Network and the algorithms to simulate every scenario analyzed. We are going to show how these algorithms work with the decision-make of the investors. We also discuss the role of the hub of the system (SFN) over the wealth distribution. Eventually, we will  present some results from our simulation where we explain the effect of both strategies over the wealth distribution of the investors and their profitable return. In the conclusion section, we state the impressive result we have obtained when comparing the return rate of the anti-imitators with the imitators one.  

\section{Methodology}

In a previous work \cite{Stefan}, we presented a Hybrid Cellular Automata (HCA) model and studied  four different kinds of networks (Regular, Random Conservative, Random Non-Conservative and Small World Network) and their influence in the Stock Market Index oscillations. This Agent-Based Model  consists of the HCA  which is able to apply a Monte Carlo process. Each node is as an investor (agent) having a psychological profile (Imitator, Anti-Imitator and Random Trader), a state (buying, holding, selling) and a number of links (connections). By using a complex network, we construct the trust neighborhood of the investors which is given by the connections that each of them has through the SFN. The states and psychological profiles are placed at random into the trust network.

Being an imitator means that an investor will perform the same state as the majority in his neighborhood. Anti-Imitator profile means that an investor will perform the same state as the minority of his neighborhood. Random Trader profile means that an investor will take his decision randomly.  

Thus, by setting up this new algorithm, we are going to give the system a dynamic process where all the investors still keep their psychological behavior, but they will not necessarily rely on their trust neighborhood. They are also going to take into account a technical analysis based on momentum technique \cite{Murphy,Pring,Steven,Rapisarda}. Accordingly, we have as the first strategy (strategy-1) the trust neighborhood and as the second strategy (strategy-2) the technical analysis. The model is built by setting the trust network of the investors (strategy-1) in order to estabilish an initial time series (100 time steps). The index experiences a fluctuation due to the dynamics of the system during these first $100^{th}$ time steps. After that, the algorithm starts performing a technical analysis over the time series which was just constructed. In this way, from the $101^{th}$ time step, this new algorithm runs both two strategies. 

Since we are using a SFN, we know that there are some nodes without any links. In this case, these nodes (investors) are going to behave as stubborn ones when considering the strategy-1 \cite{Stefan}. Therefore, those investors who do not have any connections might change their states by considering  only the technical analysis (strategy-2). We can see that at every time step all the investors will look at their neighborhood (except the ones who do not have any links) and will analyze the trend of the index. The HCA, then, will give the rules for choosing a new state considering the psychological profile of the investors. By doing so, every node (agent) will be updated synchronously. At every time step we compute the index that was, initially, set to $100$ as an initial condition.  The algorithm starts running,  updating the index considering the number of buyers and sellers. If there is more buying than selling, the net balance is going to be positive and it makes the index increase; on the other hand, if there is more selling, the net balance is negative and it  makes the index decrease \cite{Stefan}.

The implementation of a particular technical analysis, which is used in real markets to study the trend of the index, follows the  momentum (MOM) technique \cite{Steven,Murphy,Rapisarda}. The value of the momentum, $M_{\tau}(t)$, will be given by the following three equations:
\begin{equation}\label{mom}
  \begin{split}
    M_{1}(t)=I(t-1)-I(t-2)\\
    M_{2}(t)=I(t-2)-I(t-6)\\
    M_{3}(t)=I(t-6)-I(t-11)
  \end{split}
\end{equation}

where $M,t,\tau \quad \mbox{and}\quad I$ stand for Momentum, Time, Time-Lag and Index, respectively. 

In our model, this process (MOM) consists of computing the difference between the value of the index of the three different windows: $M_1(t), M_2(t)\;\mbox{and}\; M_3(t)$. We, then, have three different measures in a given trading interval (days - time steps, for instance). From the equation \ref{mom}, computing $M_{\tau}(t)$, we can assume either positive or negative values, which can be seen as a expectation of the index of going up or down, respectively. Consequently, the traders expect an increase of the index for the next time step when $M_{\tau}(t)>0$, on the other hand, if $M_{\tau}(t)<0$ they expect a decrease. Based on the technical analysis, the simulations will consider the value of the momentum as being the difference of the values of the indices  at different time-lags (windows).  

We can think about the steepness of the slope of the index: As $M_{\tau}(t)$ can assume either negative or positive values, we measure how positive or negative it is for each given $\tau$. It means that the higher a positive/negative value of the ($M_{\tau}(t)$) is the steeper the index becomes. Thus, the value of the $M_{\tau}(t)$  obtained for each time-lag proposed is going to be used to set a probability weight on the process of buying, holding or selling, as shown on the Table \ref{prob}. This table shows several patterns which depends on how steep the slope is, either increasing or decreasing. The probabilities that have been applied to this process are heuristic weights based on the combination of the three slopes, $(\tau=1,2,3)$, which are going to be interpreted as a trend of the index. 

\begin{figure*}[hbt]
  
  \centering
   {
   \includegraphics[scale=0.25,width=0.3\linewidth]{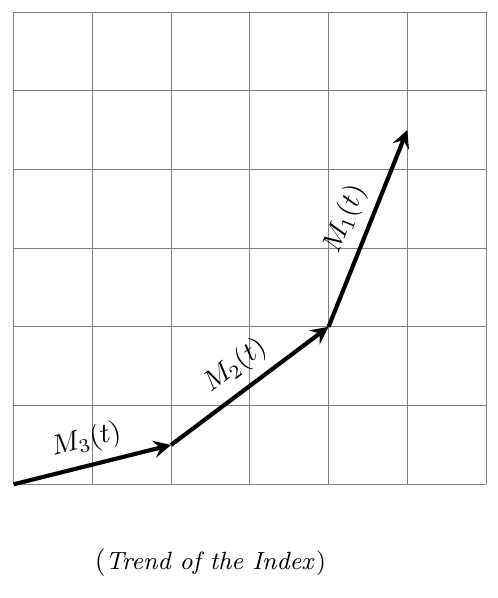}%
   \quad
    \includegraphics[scale=0.9,width=0.5\linewidth]{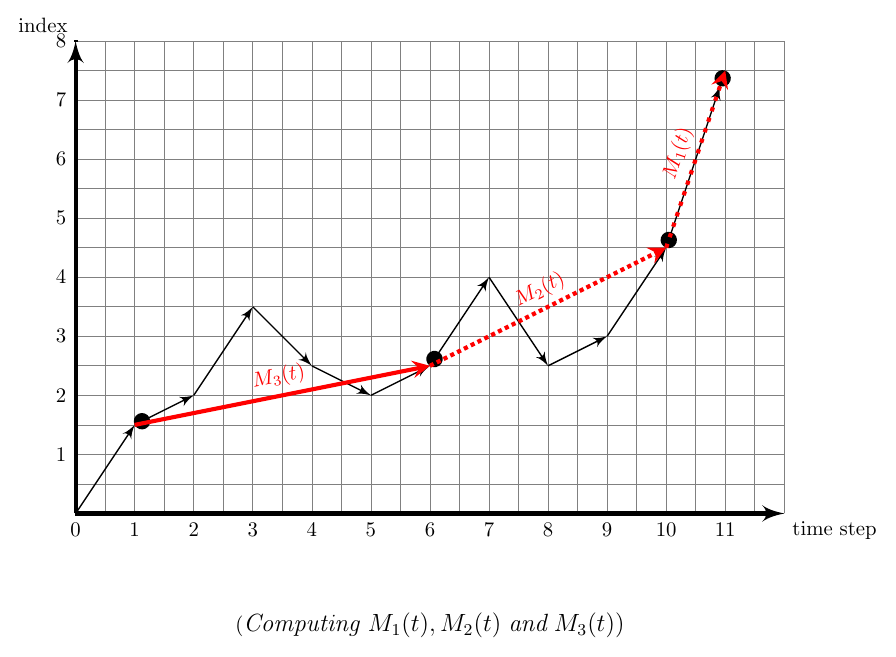}
   }
  \caption{Left: Trend of the Index where $M_{1}(t)>M_{2}(t)>M _{3}(t)>0$ to which can be applied the probabilities from the $2^{nd}$ row of the Table \ref{trend}; Right: Picture shows how the values of $M_{1}(t), M_{2}(t)\; \mbox{and}\;M _{3}(t)$ are computed.\label{slope}}%
\end{figure*}

Let us consider, for instance, if the slope gets steeper with smaller time-lag, see Figure \ref{slope}, it is interpreted as the stock price is getting more expensive. This scenario sounds like the investors should buy stocks, once there is a high probability that the stock price will increase in the next time step. In this scenario we have set a probability weight  of $(1, 0, 0)$ (although this example is a deterministic decision meaning that every investor will buy stocks)~-~(Case-1)~-~for (buying, holding, selling) respectively, see row 2 on the Tables \ref{trend} and \ref{prob}. Based on this table \ref{prob} and applying a Monte Carlo process, we set a stochastic process in order to determine whether the investor is going to decide to buy, hold or sell stocks.

From the neighborhood and technical analysis strategies, an investor can experience either the same decision-make or different one to buy, hold or sell his stocks, see Figure \ref{algor} and Algorithm-1. In order to study the system dynamics, we have implemented a stochastic process where we will consider some different probabilities from a normal distribution. It will be set the probability to be: $1\%, 5\%, 30\%, 50\%, 70\%, 95\% \,$ and $\, 99\% \,$ to follow the decision-make from the trend of the index (MOM). Moreover, it will be taken into account the behavior profile of the investors in order to follow the decision-make given from the MOM. Considering a scenario where an investor has an imitator profile, one will perform one's decision as following: he/she looks at his/her trust neighborhood and then performs the technical analysis: if one gets from both strategies the same decision-make, one will just follow it. If the strategies show a different decision-make, a stochastic process is set to decide which one will be taken. For example: strategy-1 (neighborhood) comes up  to buy stocks; strategy-2 (technical analysis) comes up to sell stocks, thus the imitator investor will then have a probability, as those stated before, to decide if he is going to be influenced by the external factors (technical analysis) or by his neighborhood (trust network). 

An interesting scenario occurs when an investor has an anti-imitator profile: he/she will always take the opposite decision which comes from the strategy-2. This situation is interesting, because we have to realize which one is the opposite for holding stocks (buy or sell?), for instance. Then, we set a stochastic process in order to decide which decision-make he/she is going to make. See Figure \ref{algor} and Algorithm-2, .

Finally, the scenario can be summarized as the following: consider that an investor has an imitator profile and 20 connections. The code shows to him how many investors are buying, holding and selling stocks. Then, at the next time step, he will make a decision to either buy, hold or sell depending on which the majority of his neighborhood was performing at that given time. Suppose that: buy=12, sell=7 and hold=1, the strategy-1 tells the investor to buy at the next time step. On the other hand, if the decision-make from the strategy-2 is the same as the one he already has, he will then buy stock at the next time step. But, if the strategy-2 gives him a contradictory decision, he will have a probability to follow the decision-make from the technical analysis. Considering, now, that this investor is an anti-imitator: if the decision-make from the strategy-2 is the same as the strategy-1, three situations can happen: first one - if both strategies show buy, he will then sell; second one - if they show sell, he will then buy; and the last one - if they show hold, he will decide between buy or sell. On the other hand, if the decision-make from both strategies is contradictory, a stochastic decision is set to decide which decision he is going to take: if the decisions from the strategy-1 and strategy-2 are respectively: 1) buy-sell: he is going to buy stock; 2)buy-hold or sell-hold: he is going to decide between buy and sell; 3) hold-buy or hold-sell: he is going to decide between hold and sell or hold and buy, respectively; 4) sell-buy: he is going to sell stock.

We decided to study the system after getting data from running the case-1, see Table \ref{prob}, and getting statistical results which can be seen on the Figures \ref{return} and \ref{hub}. Those results about the wealth distribution and rate of return made us to think about: what if we invert the system by changing the probabilities between the imitator and anti-imitator profiles. Then, we created the case-2, see the Table \ref{prob}. As expected, the system inverted the results, see Figures \ref{return} and \ref{hub} for the case-2, surprisingly, these results were not so strong as the case-1. We, then, set up the case-3 which brings the system to the balance, see Table \ref{prob}, and once again the results showed, statistically, the same results as we had from case-1. Finally, we set up the case-4 which is the opposite probabilities of the case-3. We showed only the results from the simulation of the case-4, see Figure \ref{hub}. Results from all these cases are discussed in the conclusion section where we state that the anti-imitator profile has an excellent rate of return and wealth distribution compared to imitators profile.

\begin{table*}[htb]
\begin{center}
\scalebox{0.8}{
\begin{tabular}[c]{|@{\hspace{1mm}}c@{\hspace{1mm}}|@{\hspace{1mm}}c@{\hspace{1mm}}|@{\hspace{1mm}}c@{\hspace{1mm}}|@{\hspace{1mm}}c@{\hspace{1mm}}|@{\hspace{1mm}}c@{\hspace{1mm}}|@{\hspace{1mm}}c@{\hspace{1mm}}|}\hline
\multicolumn{6}{|c|}{Trend of the Index} \\
\hline
ROW & $M_{1}(t)>M_{5}(t)$ & $M_{5}(t)>M_{10}(t)$ & $M_{1}(t)>0$ & $M_{5}(t)>0$ & $M_{10}(t)>0$\\\hline
A & 0 & 0 & 1 & 1 & 1\\\hline
B & 1 & 1 & 1 & 1 & 1\\\hline
C & 0 & 1 & 1 & 1 & 1\\\hline
D & 1 & 0 & 1 & 1 & 1\\\hline
E & 0 & 1 & 1 & 1 & 0\\\hline
F & 1 & 0 & 1 & 0 & 1\\\hline
G & 1 & 1 & 1 & 1 & 0\\\hline
H & 1 & 1 & 0 & 0 & 0\\\hline
I & 1 & 1 & 1 & 0 & 0\\\hline
J & 0 & 1 & 0 & 1 & 0\\\hline
K & 1 & 0 & 1 & 0 & 0\\\hline
L & 0 & 0 & 0 & 0 & 1\\\hline
M & 0 & 1 & 0 & 1 & 1\\\hline
N & 1 & 0 & 0 & 0 & 0\\\hline
O & 1 & 0 & 0 & 0 & 1\\\hline
P & 0 & 0 & 0 & 0 & 0\\\hline
Q & 0 & 0 & 0 & 1 & 1\\\hline
R & 0 & 1 & 0 & 0 & 0\\\hline
\end{tabular}
}
\end{center}
\caption[Table of Tendencies]{The header of the table: $M_{1}(t)$, $M_{5}(t)$ and $M_{10}(t)$ stand for the momentum considering the difference for 1 time-lag, 5 time-lag and 10 time-lag, respectively. The rows are filled in with the tautology (1:true; 0:false) and are used to build up the probabilities. \label{trend}}
\end{table*}

\begin{table*}[htb]
\begin{center}
\begin{tabular}[c]{|@{\hspace{1mm}}c@{\hspace{1mm}}|@{\hspace{1mm}}c@{\hspace{1mm}}|@{\hspace{1mm}}c@{\hspace{1mm}}|@{\hspace{1mm}}c@{\hspace{1mm}}||@{\hspace{1mm}}c@{\hspace{1mm}}|@{\hspace{1mm}}c@{\hspace{1mm}}|@{\hspace{1mm}}c@{\hspace{1mm}}||@{\hspace{1mm}}c@{\hspace{1mm}}|@{\hspace{1mm}}c@{\hspace{1mm}}|@{\hspace{1mm}}c@{\hspace{1mm}}|}\hline
\multicolumn{1}{|c|@{\hskip 1mm}}{} & \multicolumn{3}{c||@{\hskip 1mm}}{Case-1} & \multicolumn{3}{c||@{\hskip 1mm}}{Case-2} & \multicolumn{3}{c|}{Case-3}\\\hline
 ROW & P(Buy) & P(Hold) & P(Sell) & P(Buy) & P(Hold) & P(Sell) & P(Buy) & P(Hold) & P(Sell)\\\hline
  $P_{A}$  & 0.8 & 0.1 & 0.1 & 0.1 & 0.1 & 0.8 & 0.6 & 0.3 & 0.1\\\hline
  $P_{B}$  & 1.0 & 0.0 & 0.0 & 0.0 & 0.0 & 1.0 & 0.7 & 0.3 & 0.0\\\hline
  $P_{C}$  & 0.8 & 0.1 & 0.1 & 0.1 & 0.1 & 0.8 & 0.6 & 0.3 & 0.1\\\hline
  $P_{D}$  & 1.0 & 0.0 & 0.0 & 0.0 & 0.0 & 1.0 & 0.7 & 0.3 & 0.0\\\hline
  $P_{E}$  & 0.6 & 0.2 & 0.2 & 0.2 & 0.2 & 0.6 & 0.4 & 0.4 & 0.2\\\hline
  $P_{F}$  & 0.6 & 0.2 & 0.2 & 0.2 & 0.2 & 0.6 & 0.4 & 0.4 & 0.2\\\hline
  $P_{G}$  & 0.6 & 0.2 & 0.2 & 0.2 & 0.2 & 0.6 & 0.4 & 0.4 & 0.2\\\hline
  $P_{H}$  & 0.1 & 0.1 & 0.8 & 0.8 & 0.1 & 0.1 & 0.1 & 0.3 & 0.6\\\hline
  $P_{I}$  & 1.0 & 0.0 & 0.0 & 0.0 & 0.0 & 1.0 & 0.7 & 0.3 & 0.0\\\hline
  $P_{J}$  & 0.2 & 0.2 & 0.6 & 0.6 & 0.2 & 0.2 & 0.2 & 0.4 & 0.4\\\hline
  $P_{K}$  & 1.0 & 0.0 & 0.0 & 0.0 & 0.0 & 1.0 & 0.7 & 0.3 & 0.0\\\hline
  $P_{L}$  & 0.2 & 0.2 & 0.6 & 0.6 & 0.2 & 0.2 & 0.2 & 0.4 & 0.4\\\hline
  $P_{M}$  & 0.0 & 0.0 & 1.0 & 1.0 & 0.0 & 0.0 & 0.0 & 0.3 & 0.7\\\hline
  $P_{N}$  & 0.1 & 0.1 & 0.8 & 0.8 & 0.1 & 0.1 & 0.1 & 0.3 & 0.6\\\hline
  $P_{O}$  & 0.2 & 0.2 & 0.6 & 0.6 & 0.2 & 0.2 & 0.2 & 0.4 & 0.4\\\hline
  $P_{P}$  & 0.0 & 0.0 & 1.0 & 1.0 & 0.0 & 0.0 & 0.0 & 0.3 & 0.7\\\hline
  $P_{Q}$  & 0.0 & 0.0 & 1.0 & 1.0 & 0.0 & 0.0 & 0.0 & 0.3 & 0.7\\\hline
  $P_{R}$  & 0.0 & 0.0 & 1.0 & 1.0 & 0.0 & 0.0 & 0.0 & 0.3 & 0.7\\\hline
  SUM & $\Sigma$8.2 & $\Sigma$1.6 & $\Sigma$8.2 & $\Sigma$8.2 & $\Sigma$1.6 & $\Sigma$8.2 & $\Sigma$6.0 & $\Sigma$6.0 & $\Sigma$6.0\\\hline
\end{tabular}
\end{center}
\caption[Table of Probabilities]{The header of the table: Case-1 - We follow the tendency of the index of going up or down; Case-2 - We invert the tendency of the index of going up or down; Case-3 - We bring the system to the balance when the sum of those probabilities of buying, holding and selling has the same result. P(Buy), P(Hold) and P(Sell) stand for the probability given for buying, holding and selling.\label{prob}}
\end{table*}
\newpage

\subsection{Complex Network}

\begin{figure*}[hbt]
  
  \centering
   {
   \includegraphics[scale=0.4,width=0.40\linewidth]{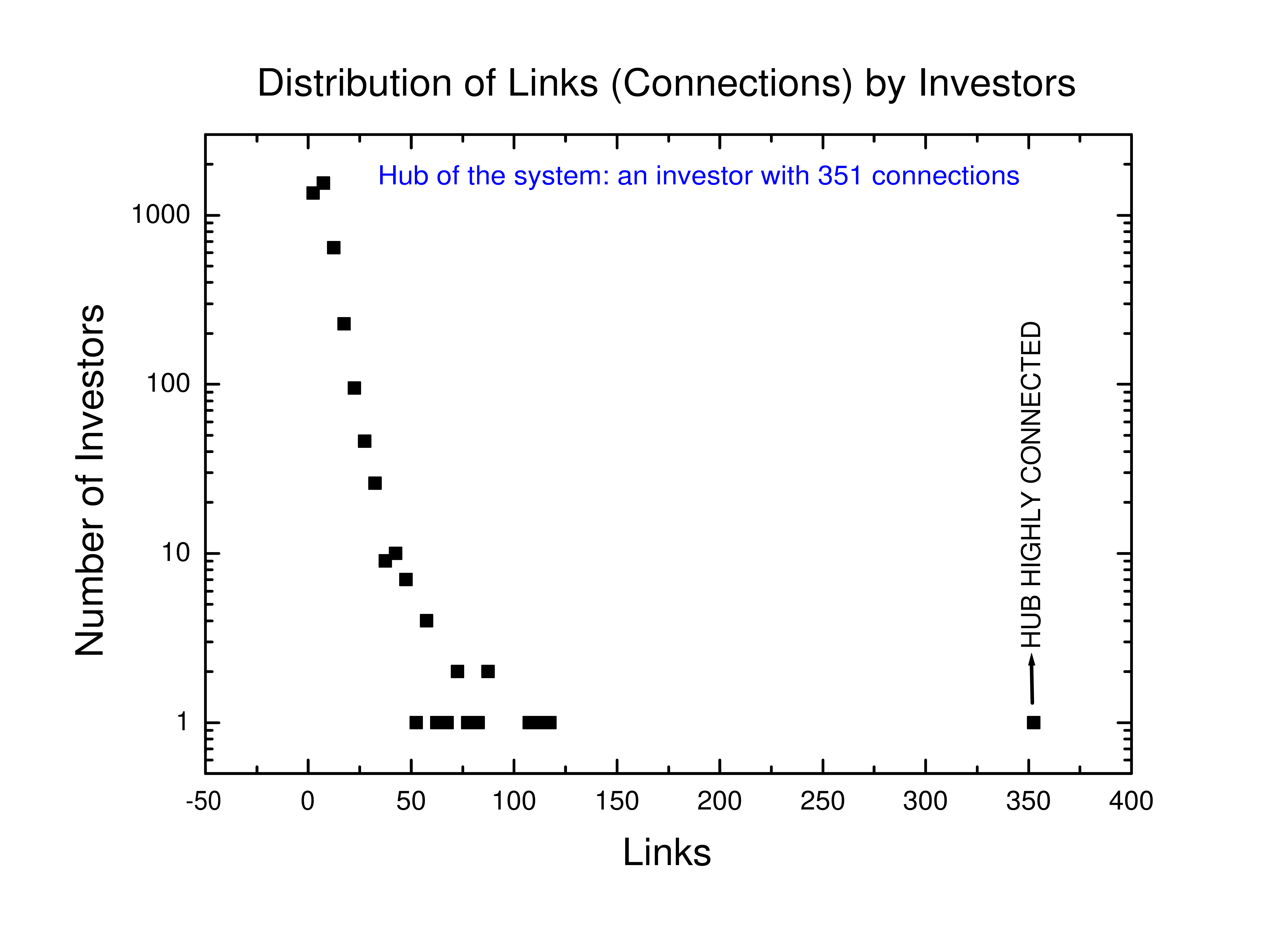}%
   \quad
    \includegraphics[scale=0.55,width=0.4\linewidth]{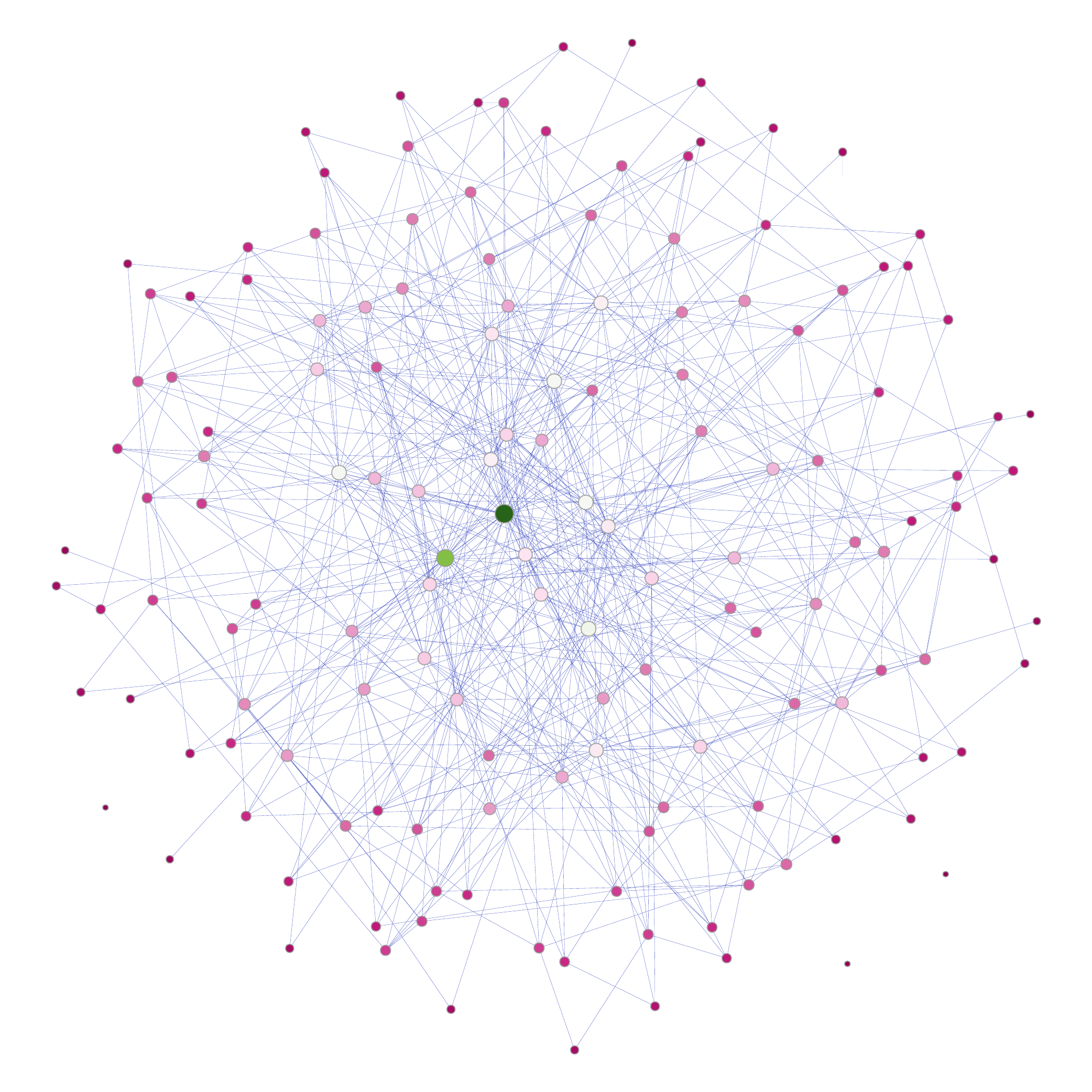}
   }
  \caption{
  Left: The graph shows that the most connected hub of the system has 351 connections followed by the one which has 116..
  Right: The Figure shows a low density where the hubs can be seen from inside out following the proportion of its connectivity degree. \label{fig:swn}}%
\end{figure*}

In this section we show the properties of the trust network which connects the investors. We set a matrix whose size is $63\times63$, where the nodes represent the investors. The Figure \ref{fig:swn},below, shows an example of a trust network with SFN morphology. As expected, it happens to have hubs highly connected which can be realized by exhibiting the distribution of investors by links, see Figure \ref{fig:swn}.\\

In the case of SFN, we consider the Barabàsi–Albert algorithm \cite{Barabasi,barabasi99} to build it. Basically, the code considers a preferential attachment of the links in such way that, the greater the number of links of a node (investor), the higher the probability of a new node to be connected to it: the rich gets richer! Thus, we are able to generate SFN up to $N$ nodes and $8N$ links, whose distribution by node follows a power law, $ \mathcal{N} (\ell) \sim \ell^{\gamma}$ . The exponent measured, $\gamma$ $\sim -2.5$, agrees with the expected value for the Barabàsi–Albert model for networks of comparable sizes.\\

We can realize that this trust network (SFN) presents a power law distribution (PLD) explaining such a highly connected investor that we are calling by hub of the system. In this case the hub has $351$ connections followed by other hubs which have, roughly, less than $125$ connections. This behavior is  characterized by the PLD.

%

\section{Results}

We present in this section some results for the two combined strategies. Firstly, we show the temporal series analysis of the stock index considering  those three cases from the Table \ref{prob}. Secondly, we are going to consider some scenarios for these combined strategy-1 and strategy-2 where we are going to dicuss the wealth distribution of the investors and the rate of return based on the influence of the hub (investor) in function of his psychological profile.

\subsection{Index Oscillation}

\begin{figure*}[!ht]
   
    \includegraphics[width=0.44\linewidth]{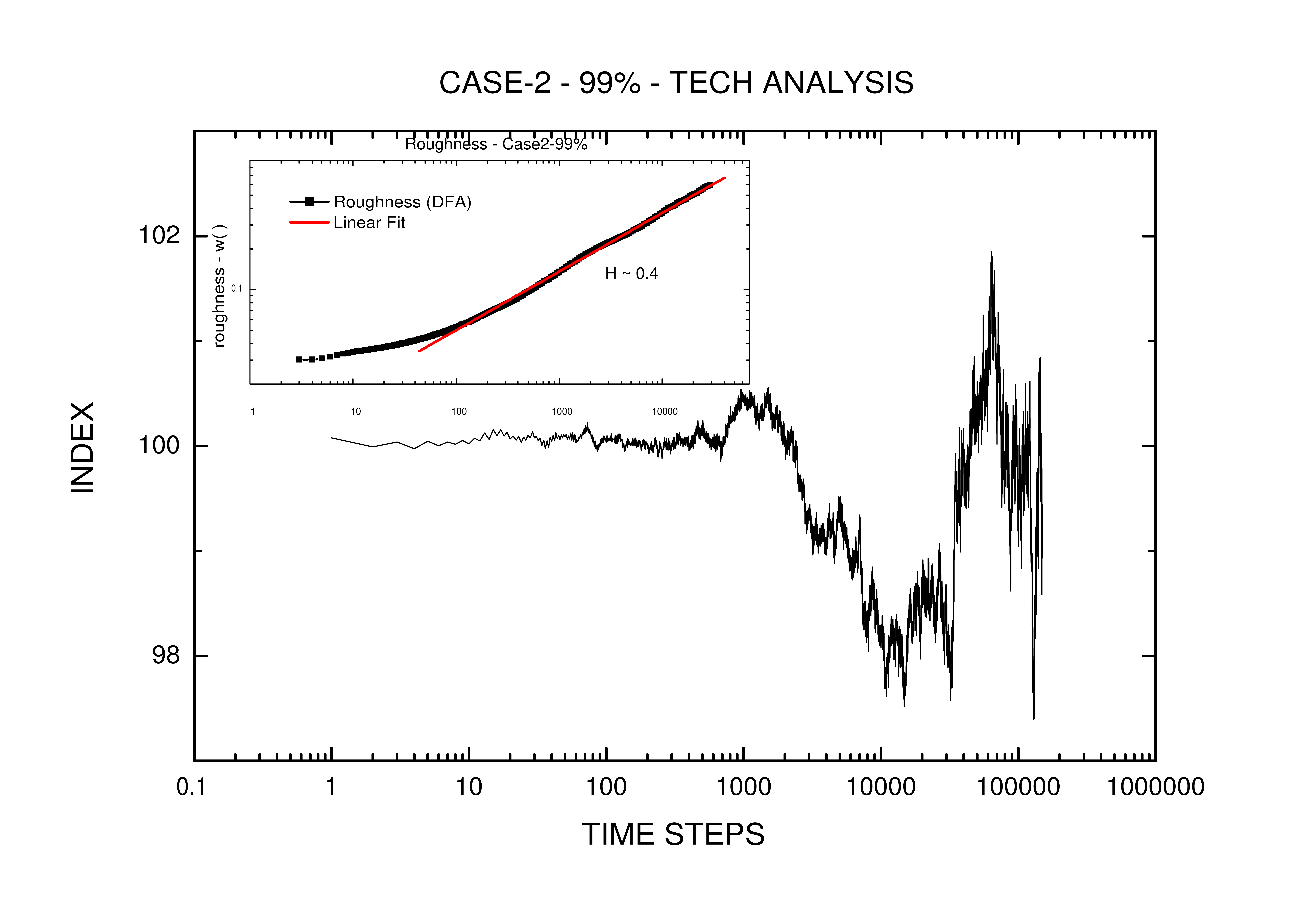}
    \includegraphics[width=0.44\linewidth]{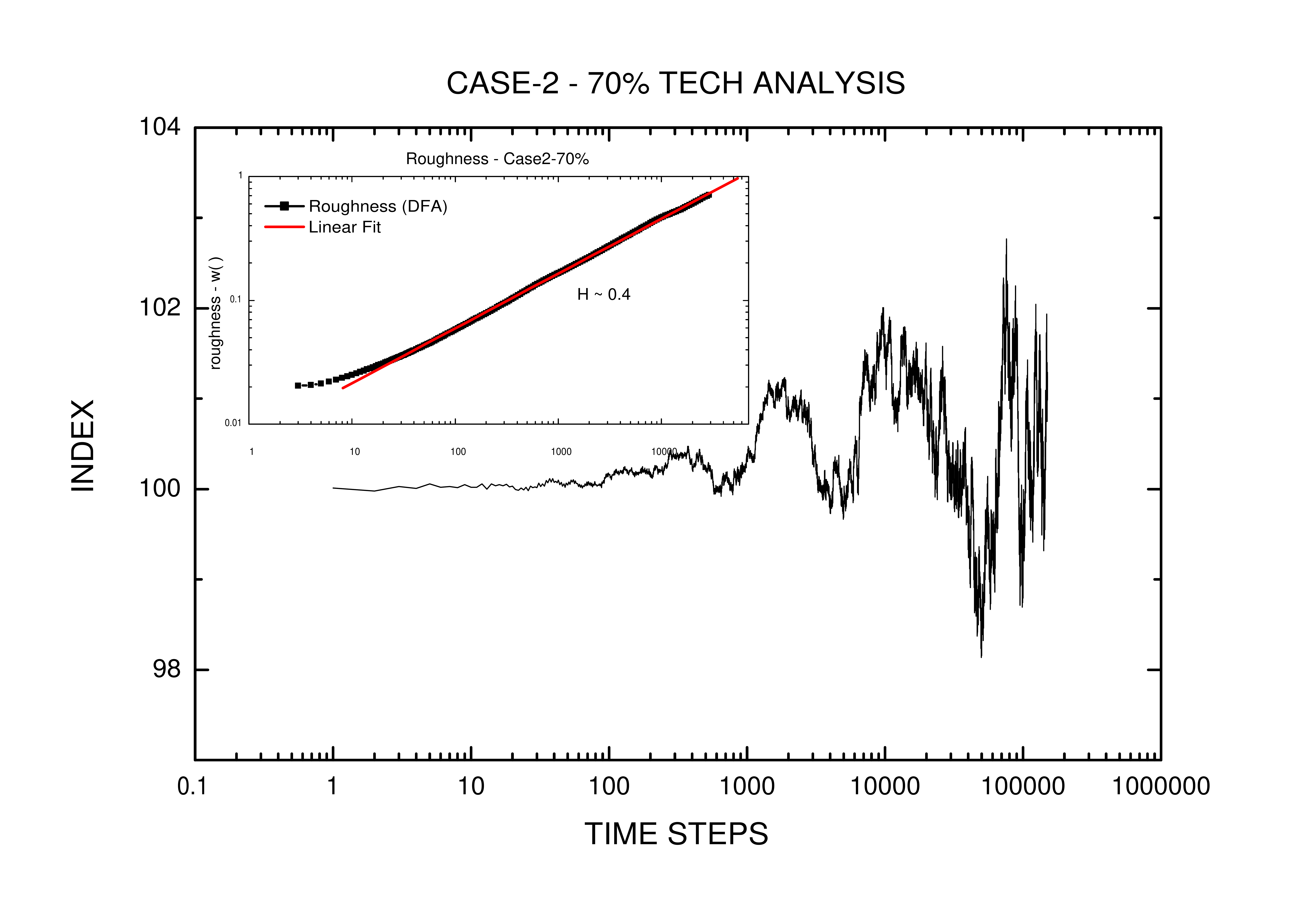}
   \caption{The figures show the index oscillations depending on the probabilities to follow the technical analysis and the specific case from the table \ref{prob} applied to set the probabilities. We verify the stochastic feature of the model by seeing the characteristic behavior of the stock market index quantified by the HE. Left: $99\%$ and Case-2; Right: $70\%$ and Case-2. \label{index}}
\end{figure*}

In this section we show some results for the Case-2, from Table \ref{prob}. In the Figure \ref{index} we  can verify that the Hurst Exponent (HE) for this case is $ H_{99\%} = 0.4375 \pm 0.001$ and $ H_{70\%} = 0.4342 \pm 0.002$. This means that there are no tendencies on the measured and it agrees with the efficient market hypothesis. The Cases $1$ and $2$ show two slopes for the HE which are going to be explored with another methodology in order to explain their role in this scenario.

\subsection{Wealth Distribution}

\begin{figure*}[htb]
   
    \includegraphics[width=0.33\linewidth]{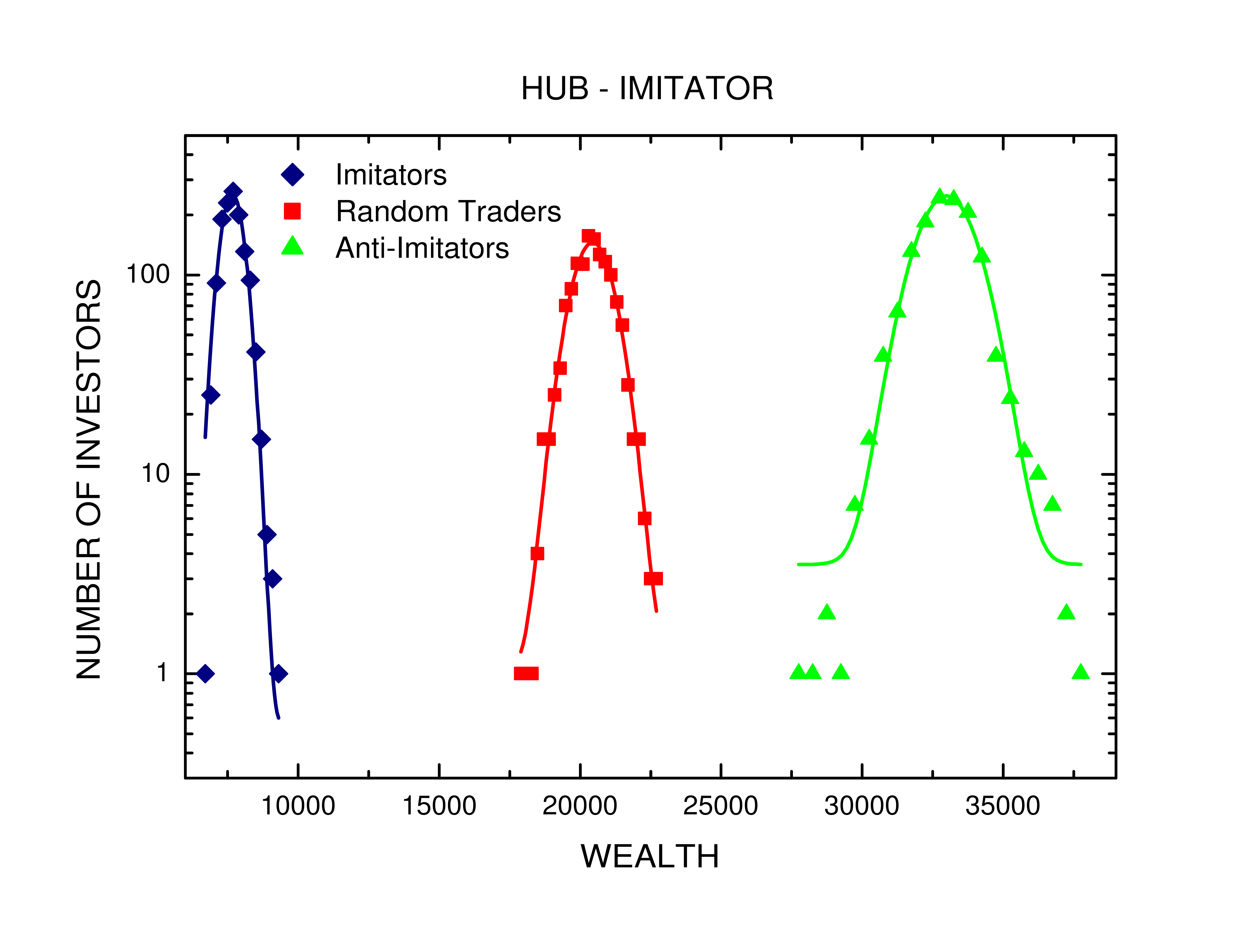}
    \includegraphics[width=0.33\linewidth]{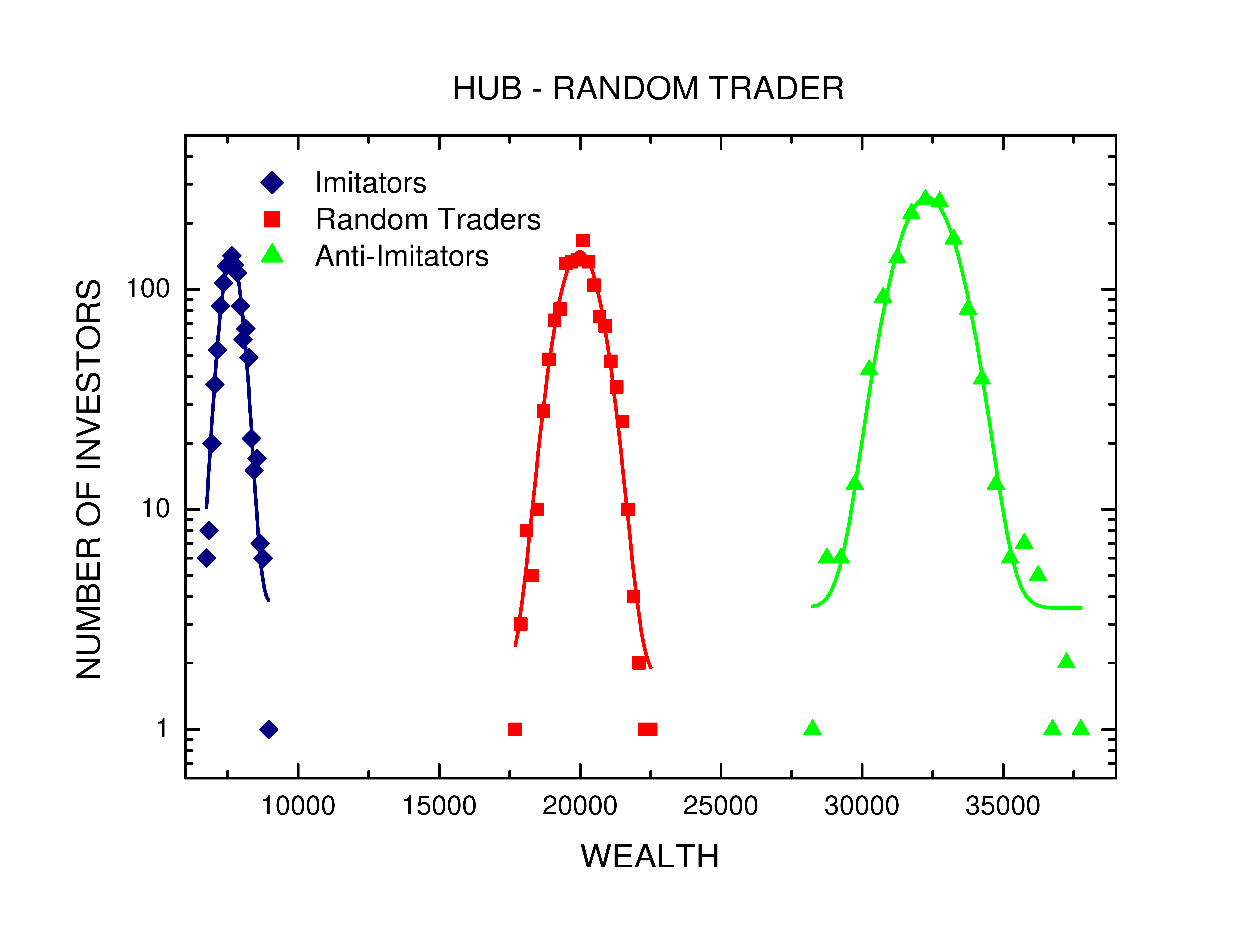}\hfill
    \includegraphics[width=0.33\linewidth]{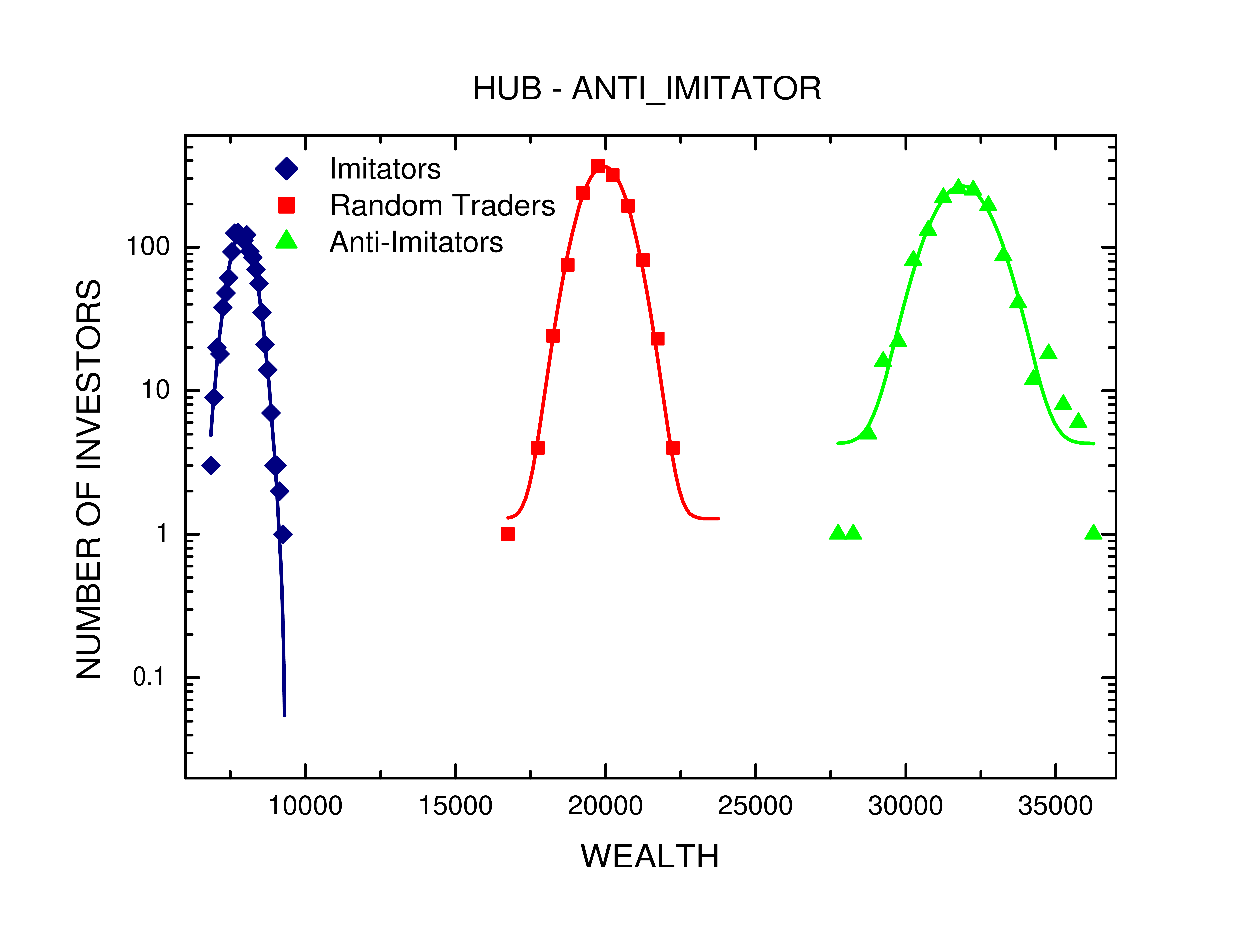}
    
    \caption{\small{Profit of the system as function of the hubs - $99\%$. Left: Hub - Imitator; Center: Hub - Random Trader; Right: Hub - Anti-imitator. For all of the three scenarios we can see straightaway that the anti-imitation profile has a remarkable perform over the stock market. The results for the whole system are:$\;\mu_{\mbox{anti}}=31878.00 \pm 23.35 \;R^{2}= 0.99345;\;\mu_{\mbox{random}}=19937.00 \pm 16.44 \;R^{2}= 0.99562; \; \mu_{\mbox{imit}}=7886.00 \pm 11.41 \;R^{2}= 0.98123 $ }\label{99}}
\end{figure*}

This section presents some results considering different probabilities to follow the technical analysis (MOM), strategy-2. From the Figure \ref{99}, we can realize how strong the MOM is by impacting on the decision-make of the investors. Pictures on the left, middle and right show the hub of the system set to be an imitator, random-trader and anti-imitator, respectively. It is notorious that, statistically, the three of them present the same results. We can see that the anti-imitation is the best psychological strategy adopted to work over this scenario as we have $\;\mu_{\mbox{anti}}=31878.00 \pm 23.35 \;R^{2}= 0.99345;\;\mu_{\mbox{random}}=19937.00 \pm 16.44 \;R^{2}= 0.99562; \; \mu_{\mbox{imit}}=7886.00 \pm 11.41 \;R^{2}= 0.98123 $. This result has been highlighted by the study of the rate of return as it is shown in the Figure \ref{return} on the next part of this section,  \ref{sec:return}, for each one of them. 

\begin{figure*}[htb]
   
    \includegraphics[width=0.33\linewidth]{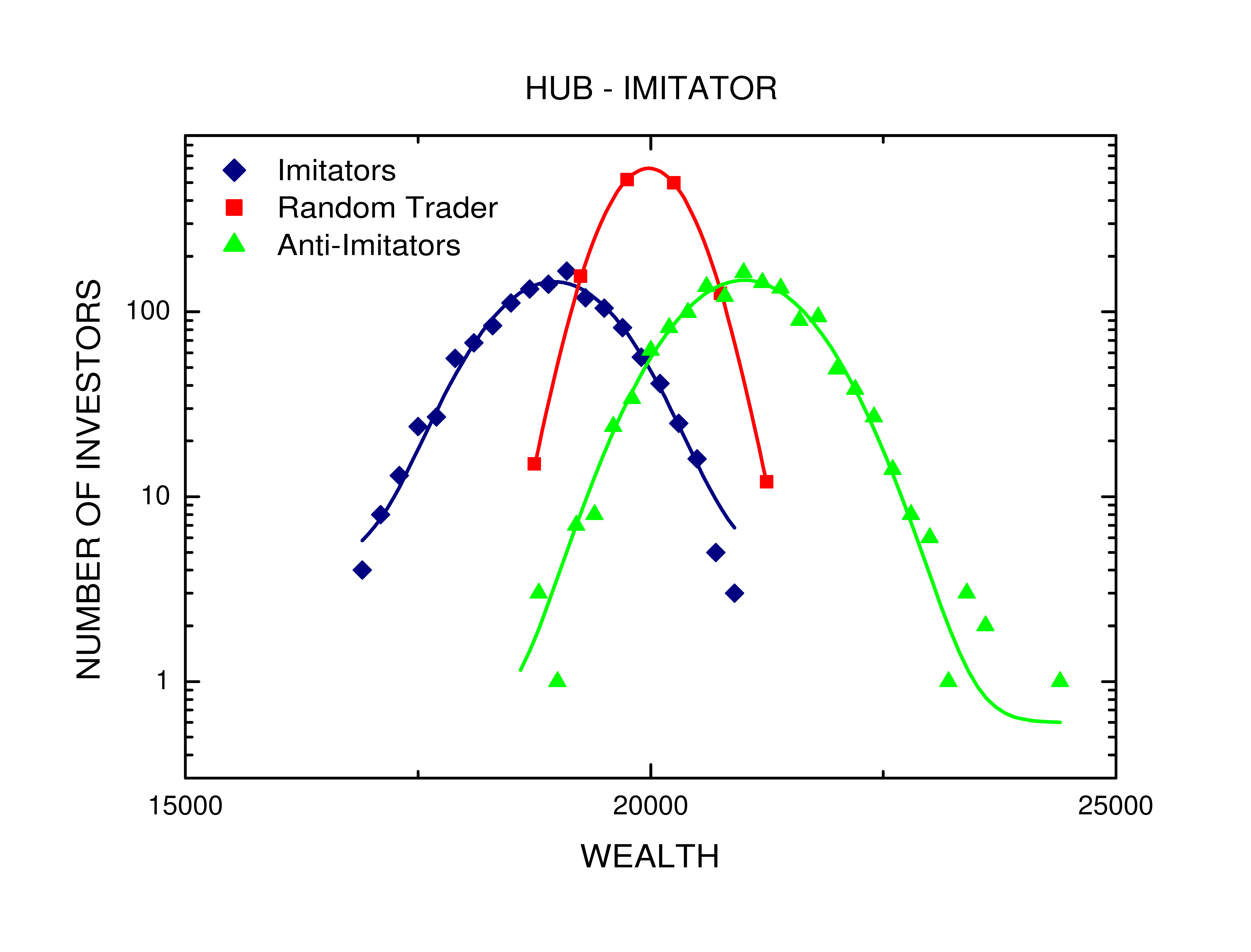}
    \includegraphics[width=0.33\linewidth]{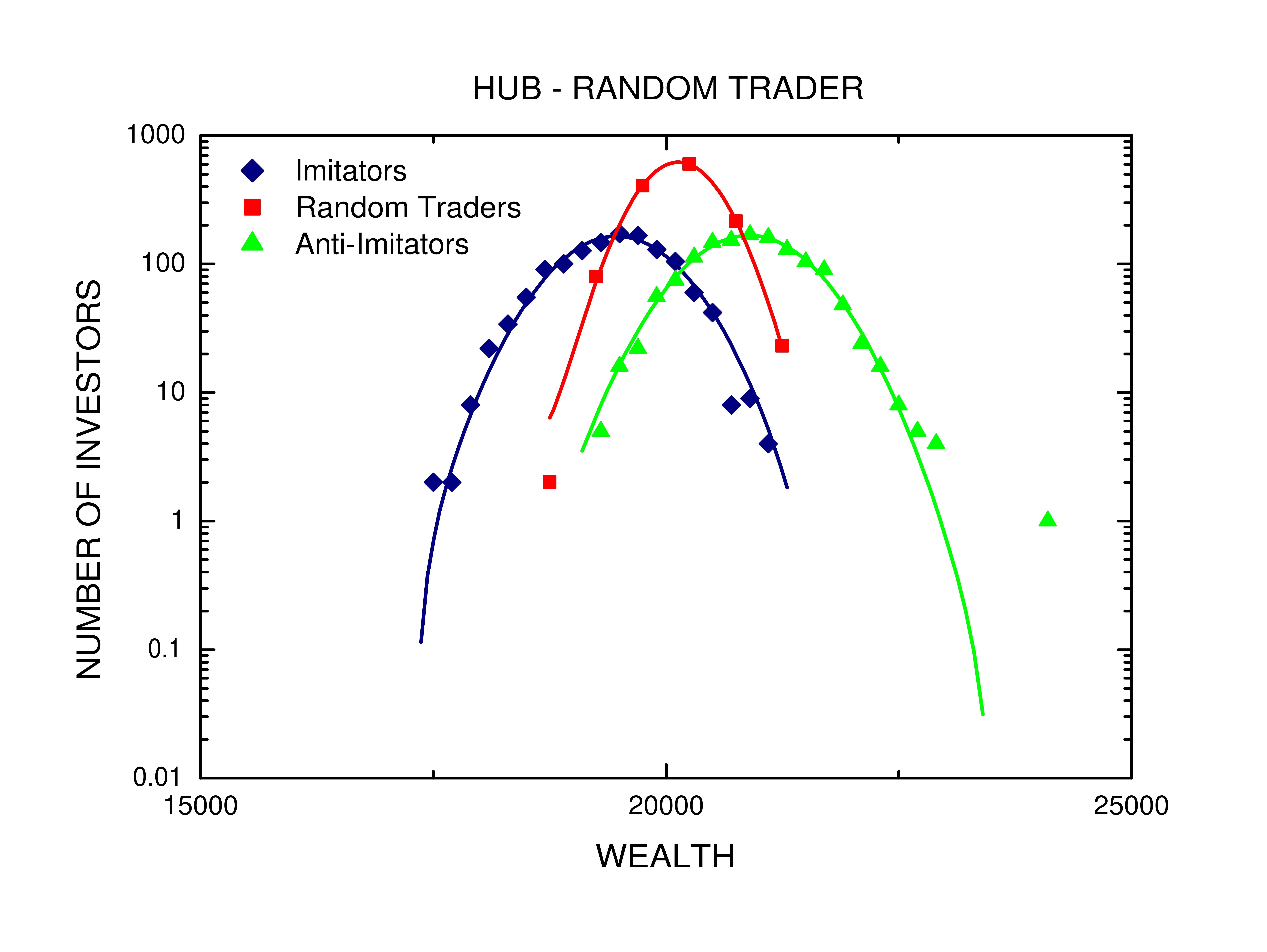}\hfill
    \includegraphics[width=0.33\linewidth]{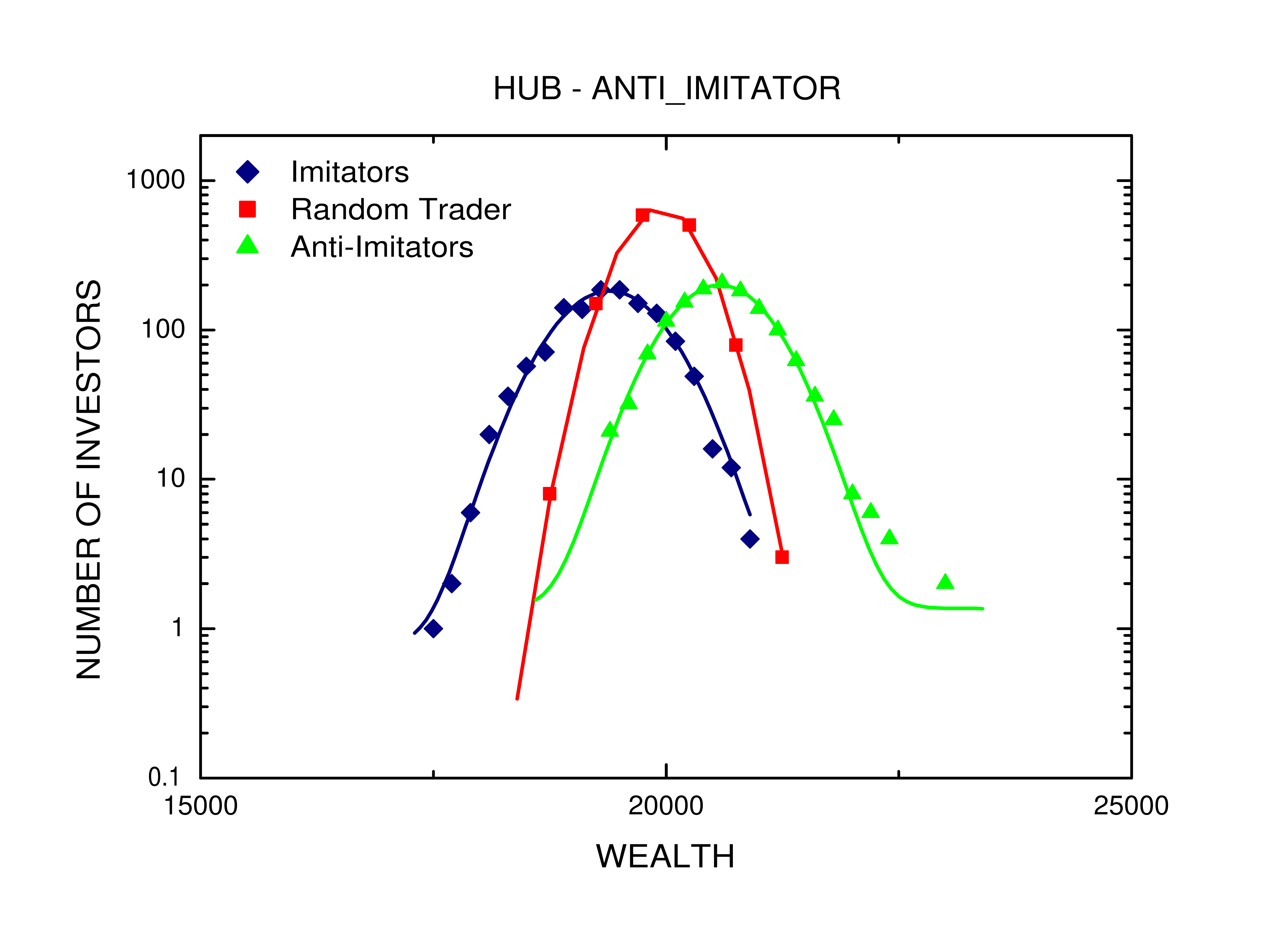}
    
    \caption{\small{Profit of the system as function of the hubs - $50\%$. Left: Hub - Imitator; Center: Hub - Random Trader; Right: Hub - Anti-imitator. Even though their profit are concentrated around $20000.00$, we can see that the anti-imitation strategy has a better perform than the others ones.The results for the whole system are:$\;\mu_{\mbox{anti}}=20577.95.00 \pm 6.64 \;R^{2}= 0.99738;\;\mu_{\mbox{random}}=20134.00 \pm 5.49 \;R^{2}= 0.99967; \; \mu_{\mbox{imit}}=18963.00 \pm 20.64 \;R^{2}= 0.98197$ }\label{50}}
\end{figure*}

\begin{figure*}[htb]
   
    \includegraphics[width=0.33\linewidth]{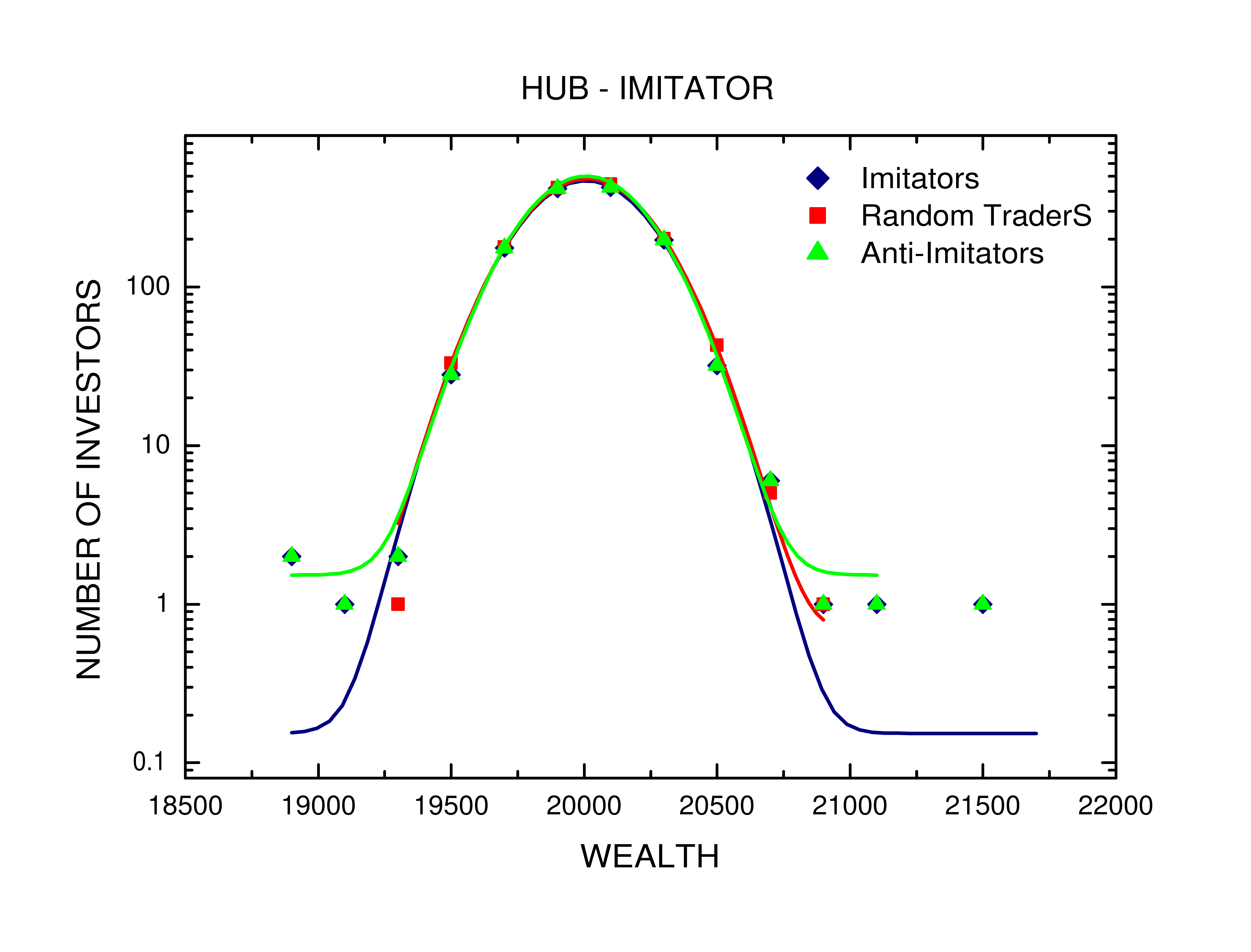}
    \includegraphics[width=0.33\linewidth]{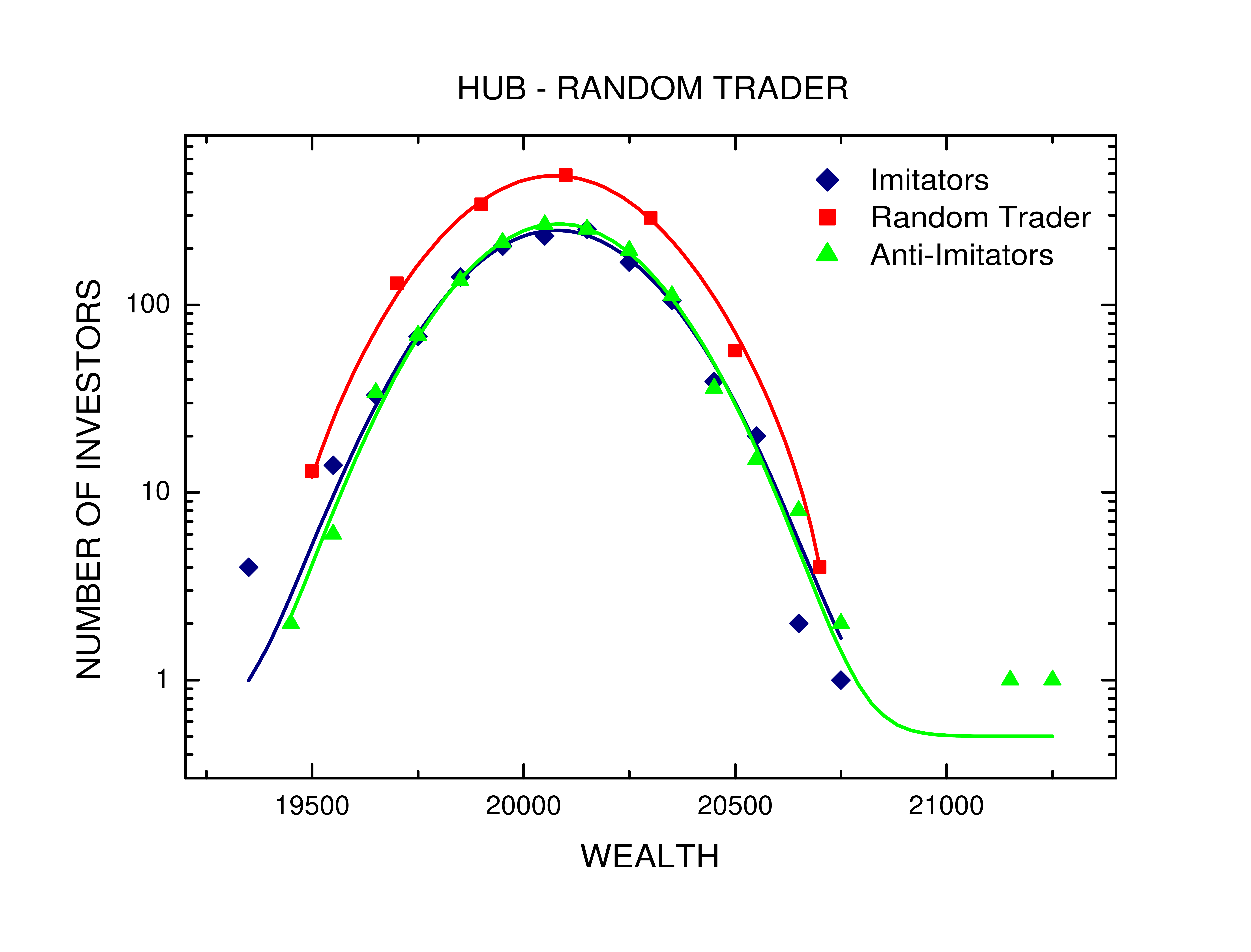}\hfill
    \includegraphics[width=0.33\linewidth]{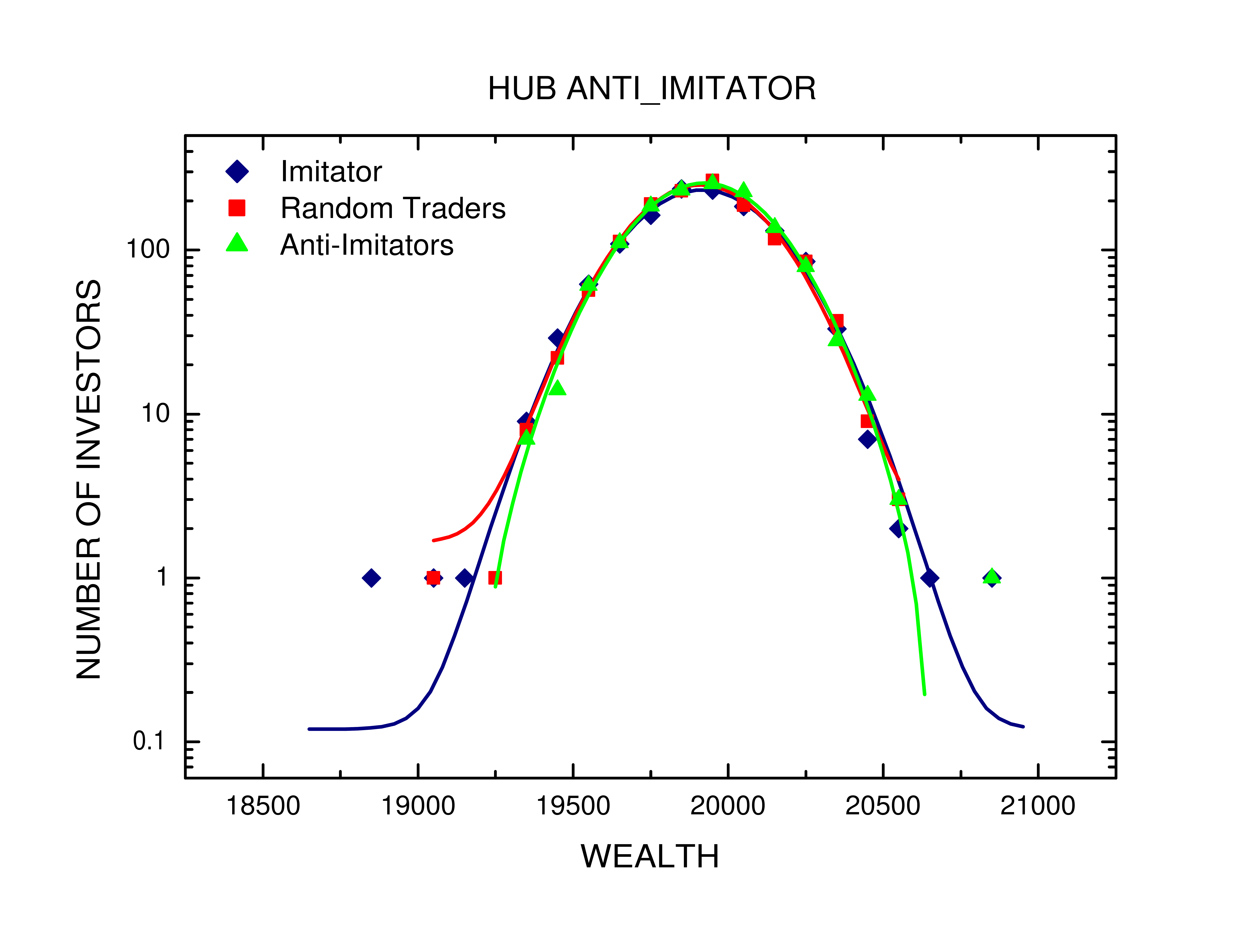}
    
    \caption{\small{Profit of the system as function of the hubs - $1\%$. Left: Hub - Imitator; Center: Hub - Random Trader; Right: Hub - Anti-imitator. They all present, statistically, the same results. Therefore, the choice for a specific psychological profile has no impact over the stock market.The results for the whole system are:$\;\mu_{\mbox{anti}}=20081.95.00 \pm 2.50 \;R^{2}= 0.9981;\;\mu_{\mbox{random}}=20073.00 \pm 7.16 \;R^{2}= 0.99644; \; \mu_{\mbox{imit}}=20077.00 \pm 5.17 \;R^{2}= 0.99325$} \label{1}}
\end{figure*}

From the Figure \ref{50}, we can realize that the anti-imitation still have a better return that the others profiles, but not as good as the previous scenario from Figure \ref{99}. Again, it does not matter if the hub of the system change his psychological profile: the system response is, statistically, the same. Now, we have results from Figure \ref{1} which show that  the wealth distribution of the investors is the same for all of them. So, it does not matter their psychological profile when setting the probability of $1\%$ to follow the MOM.
Simulations from several scenarios with different probabilities to follow the MOM technique have provided us such a remarkable result where all the investors who are anti-imitators obtained a profitable return by following strictly the MOM. Their richness does depend on the sort of the behavioral profile. We can see straightaway from the simulations that all the anti-imitators got very rich, not only the hub of the system, but all the anti-imitators. From Figure \ref{system}, we see how  the whole system displays the wealth distribution as we move the probabilities from $1\%$ to $99\%$. For each probability, we run the system with the hub set to be an imitator and we then got the average of the wealth from all the imitators in the system. Then, we did the same by setting the hub to be a random-trader and then an anti-imitator. This Figure \ref{system} gives an exactly idea of how the anti-imitator profile performing a technical analysis is far more profitable than the others two ones. 

As we increase the probability  of allowing the investors to consider the strategy-2, we realize how the anti-imitators get much richer and the imitators get very poor. Statistically, from Figures \ref{50} and \ref{1}, they all had the same return, but by increasing the probability to greater than  $50\%$, it can be seen a huge difference between wealth distribution of each psychological profile, see Figure \ref{system}. The results which come from the random-traders simulations just confirm what the literature says \cite{Rapisarda,Bouchaud,lebaron00}, therefore, it brings robustness to our model.

\subsection{Return Distribution}
\label{sec:return}

\begin{figure*}[htb]
  \begin{minipage}[t]{0.9\linewidth}
  
    \includegraphics[width=0.45\linewidth]{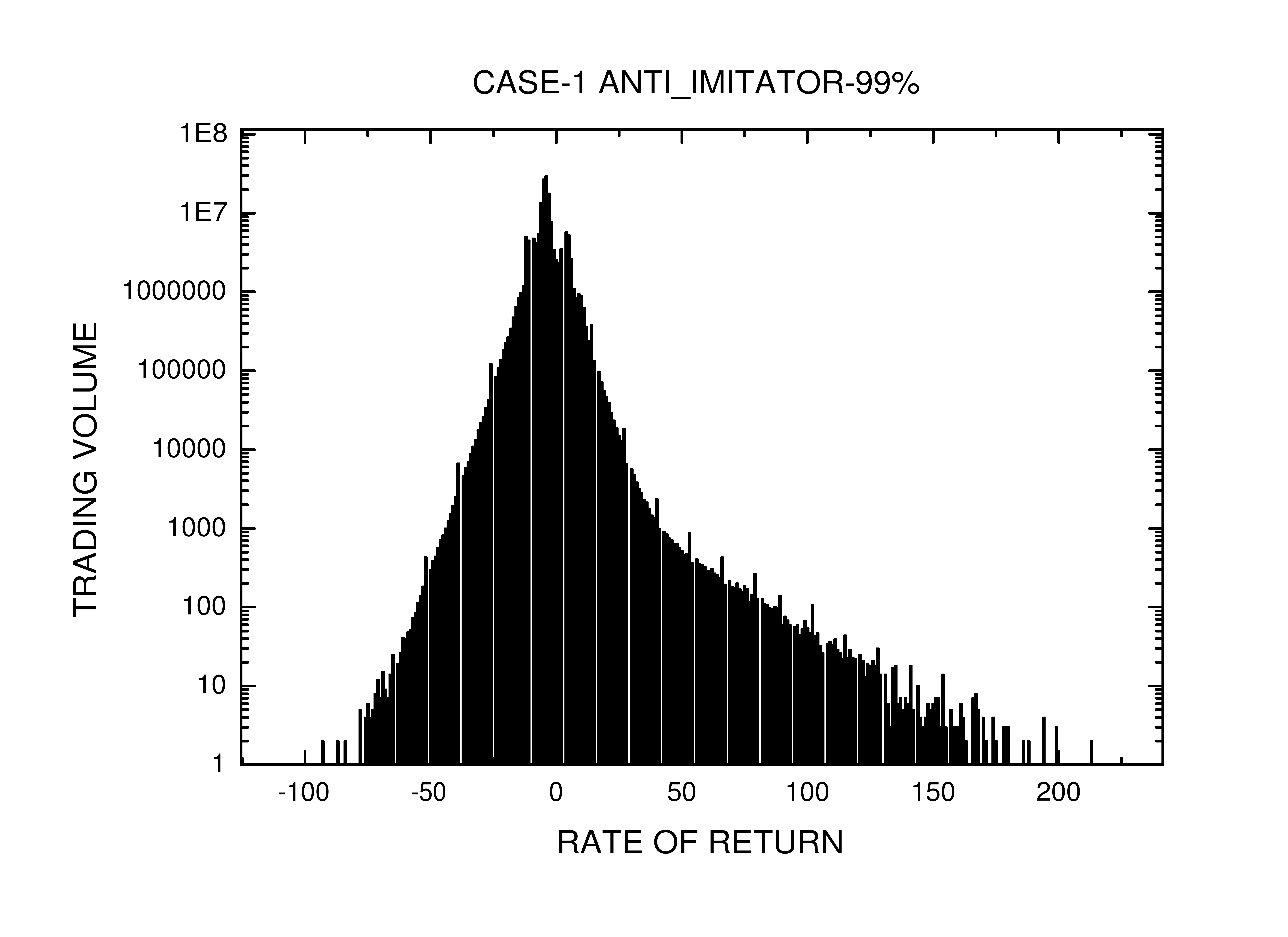}
    \includegraphics[width=0.45\linewidth]{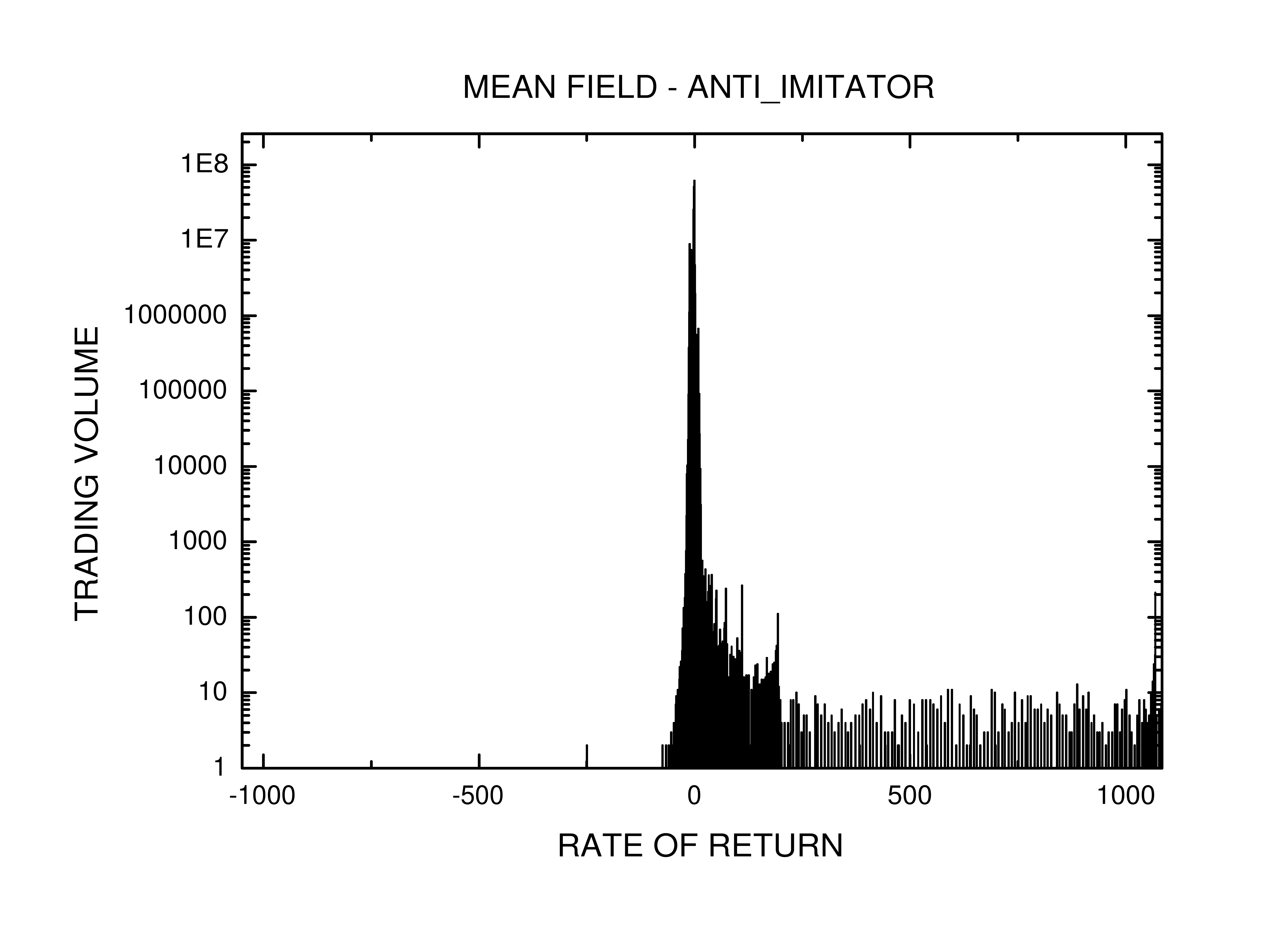}
    \includegraphics[width=0.45\linewidth]{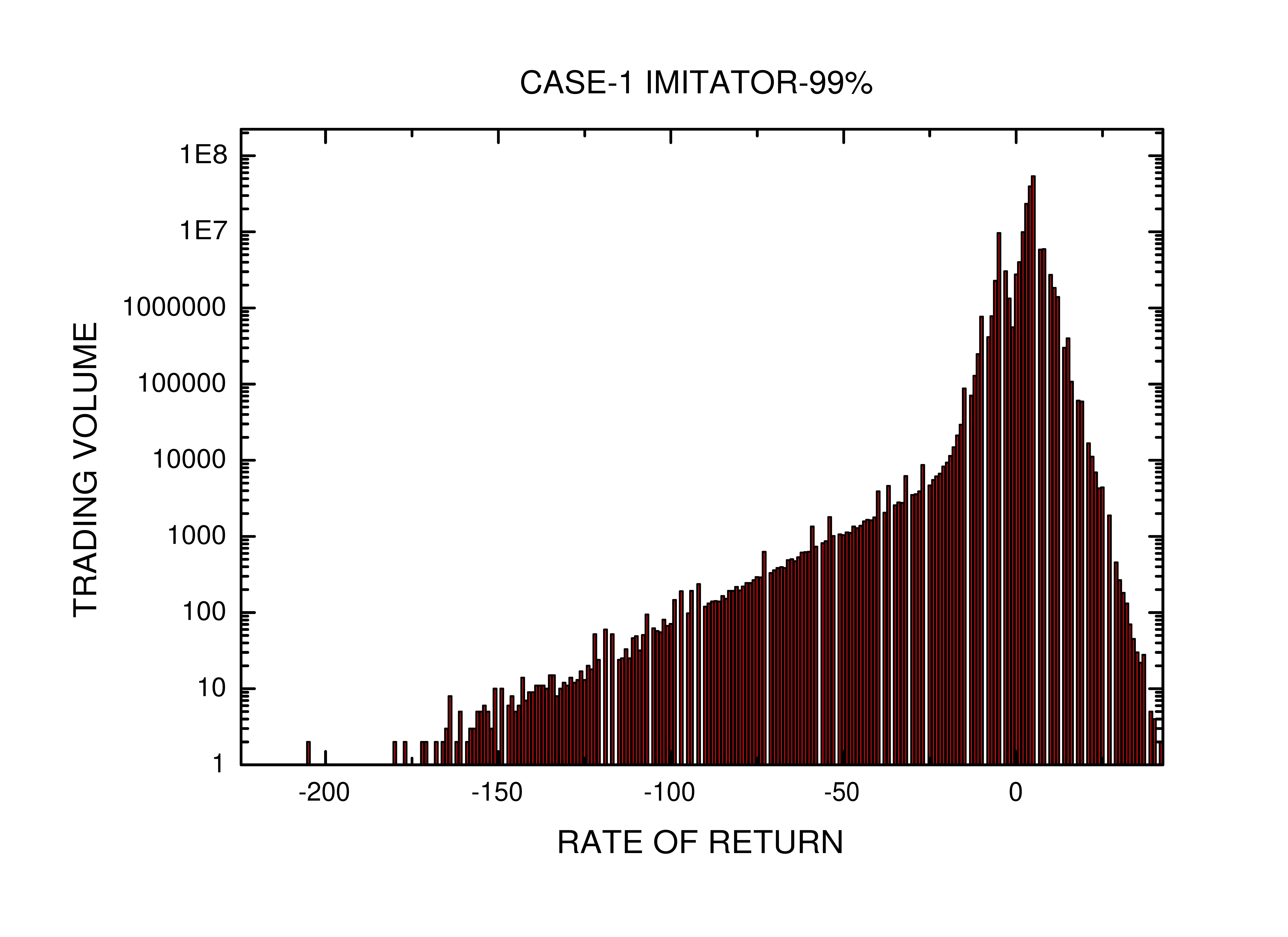}
    \includegraphics[width=0.45\linewidth]{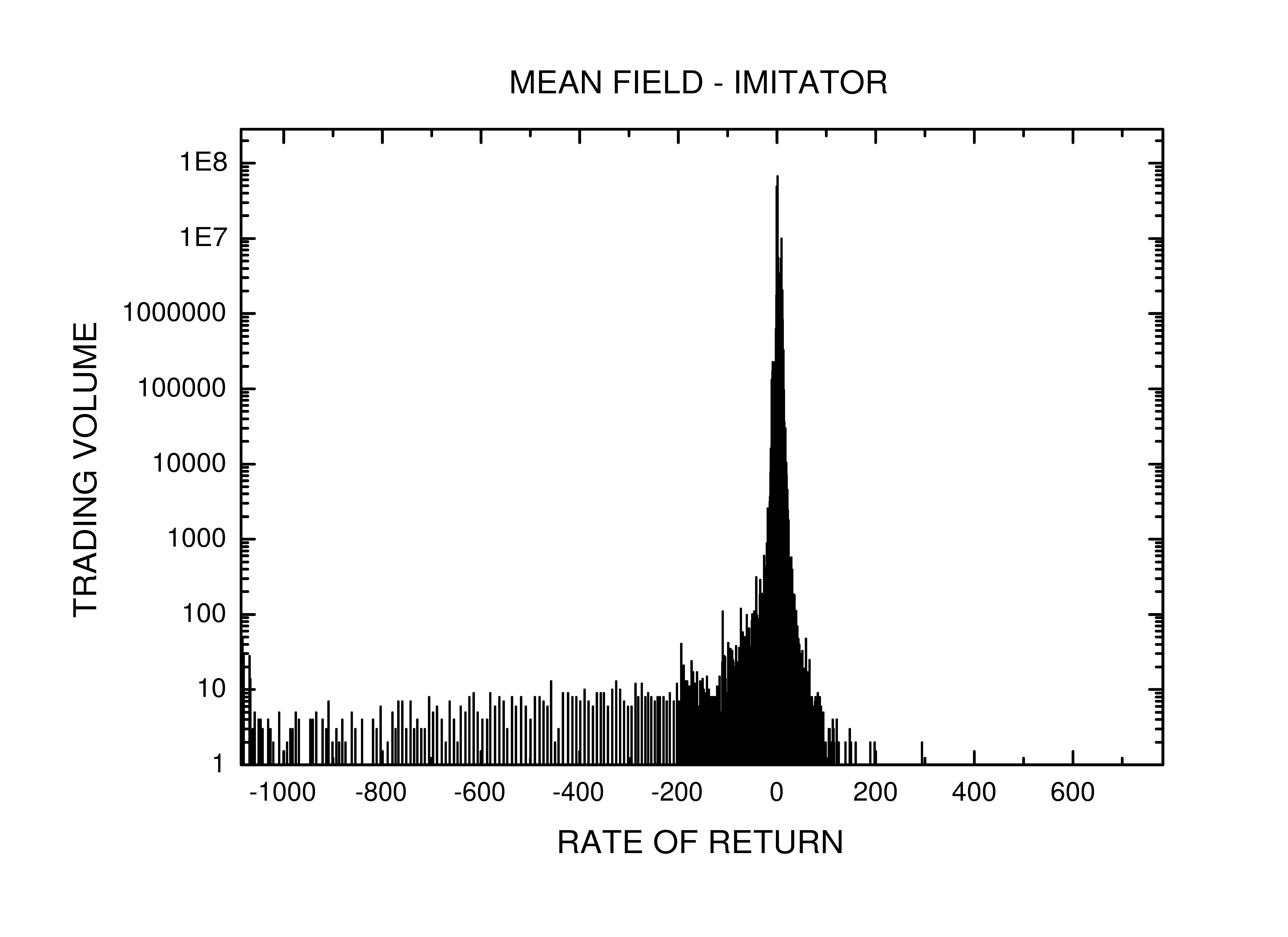}
    \includegraphics[width=0.45\linewidth]{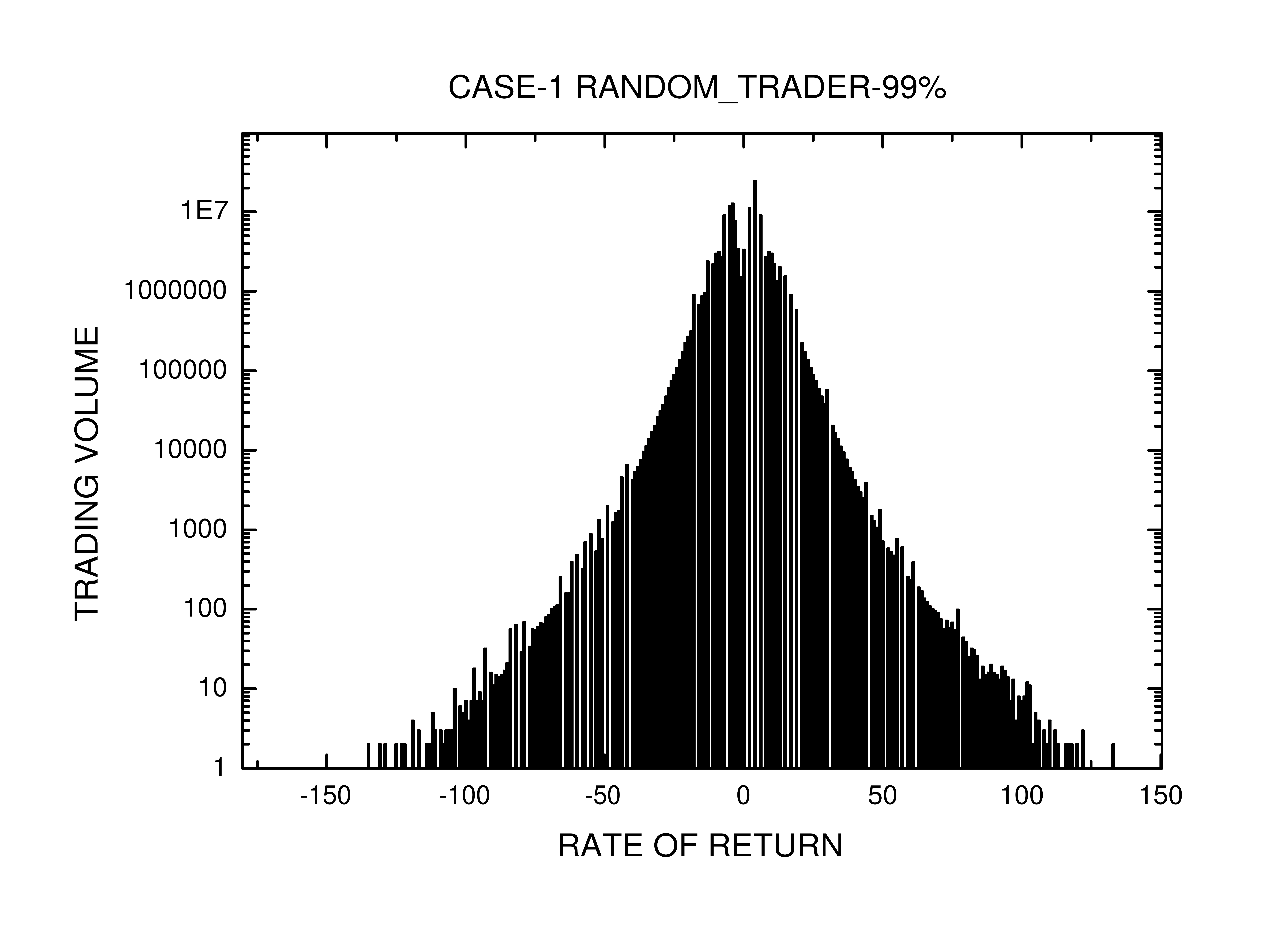}\hfill
    \includegraphics[width=0.45\linewidth]{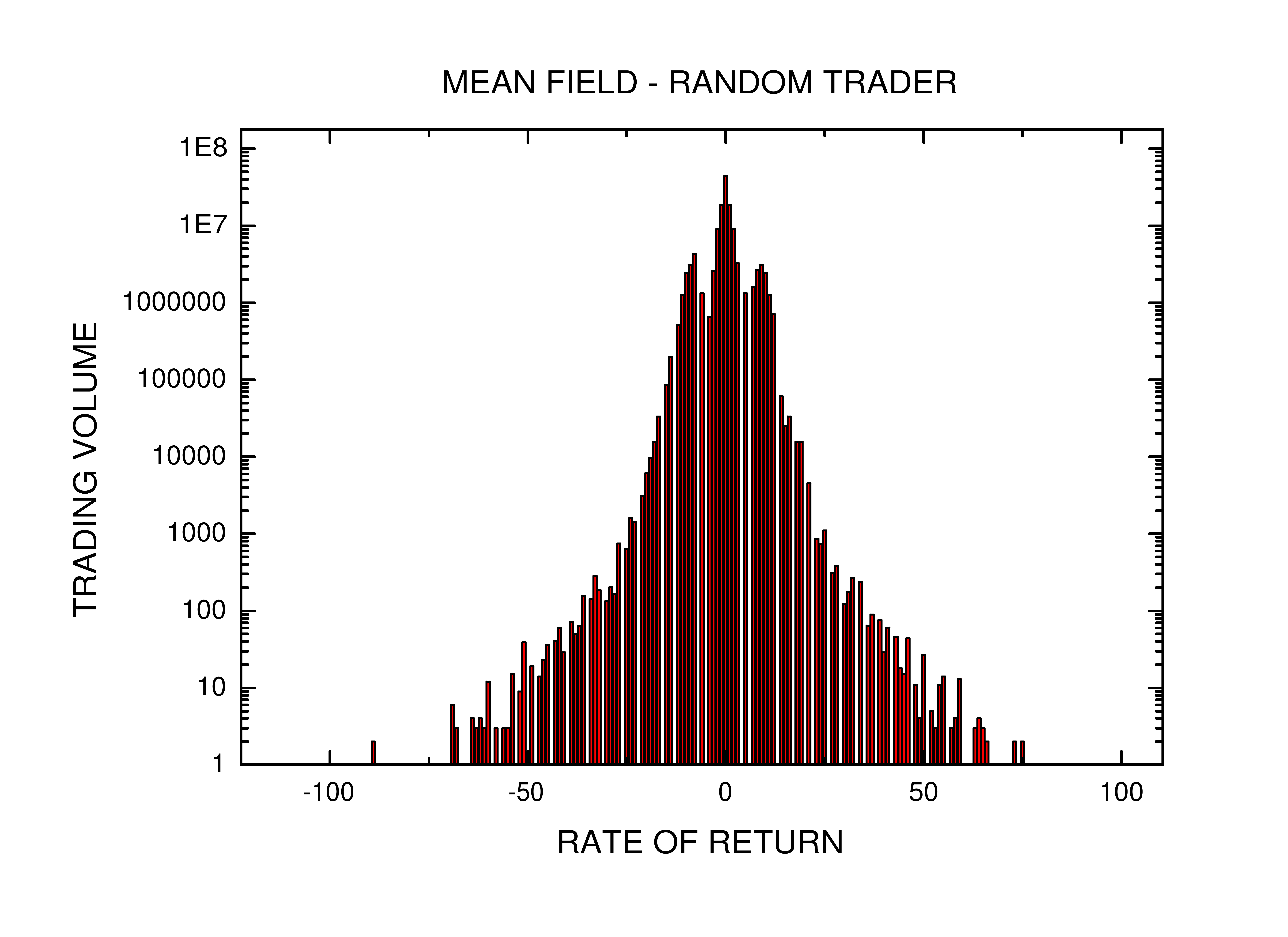}
    \caption{Rate of Return. Left side:  Applying the Case-1 from the Table \ref{prob} and a probability of $99\%$ to follow the technical analysis - Top-anti-imitators investors which is concentrated on the positive return side; Middle-imitators investors which is concentrated on the negative return side; Bottom-random-trades investors which is symmetric around the origin. Right side: the figures show, statistically, the same results as the ones shown at the left side when applying another technique to compute the decision-make.}
 \label{return}
 \end{minipage}%
\end{figure*}

\begin{figure*}[htb]
   \begin{minipage}[t]{0.9\linewidth}
    \includegraphics[width=0.45\linewidth]{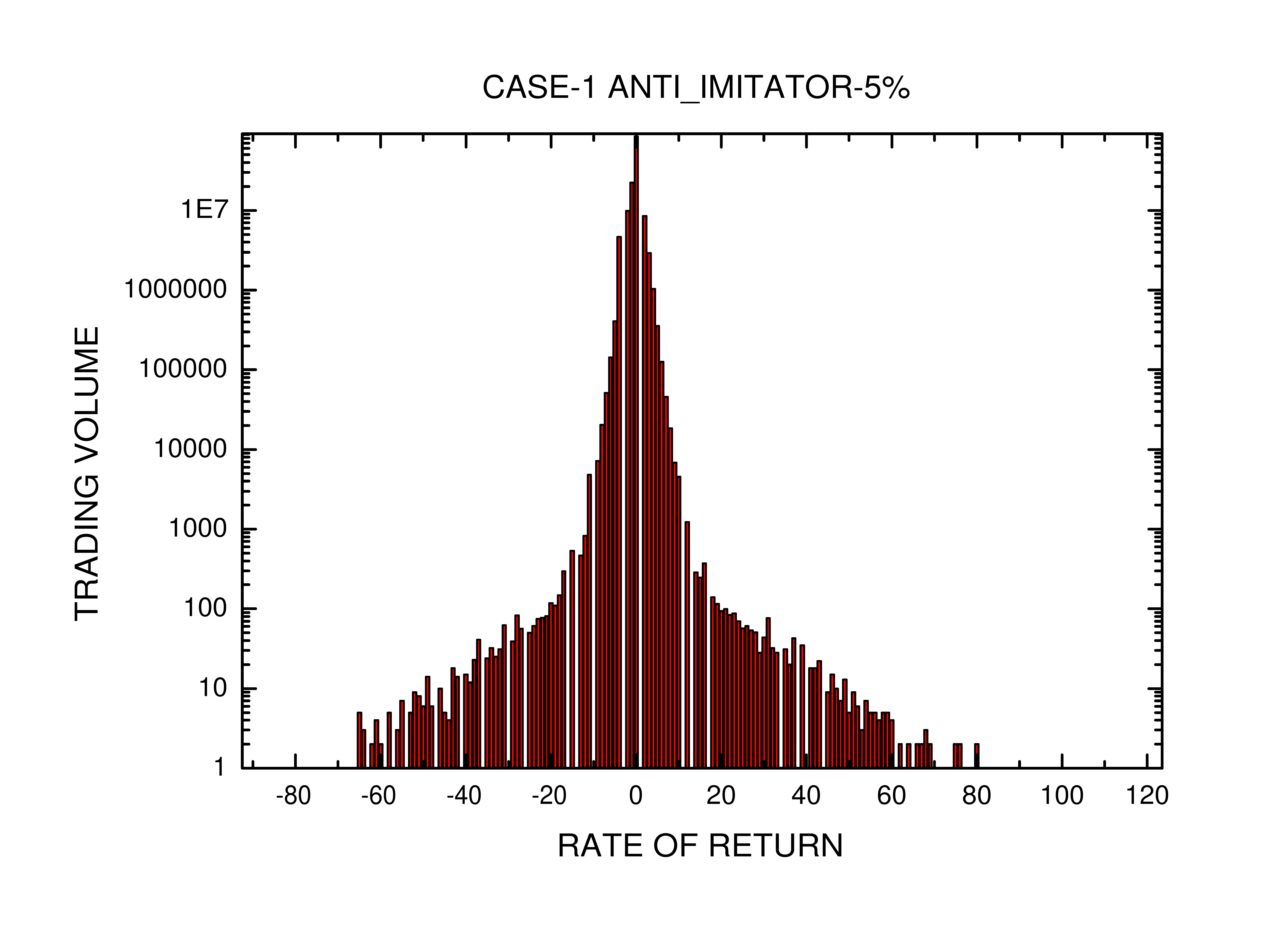}
    \includegraphics[width=0.45\linewidth]{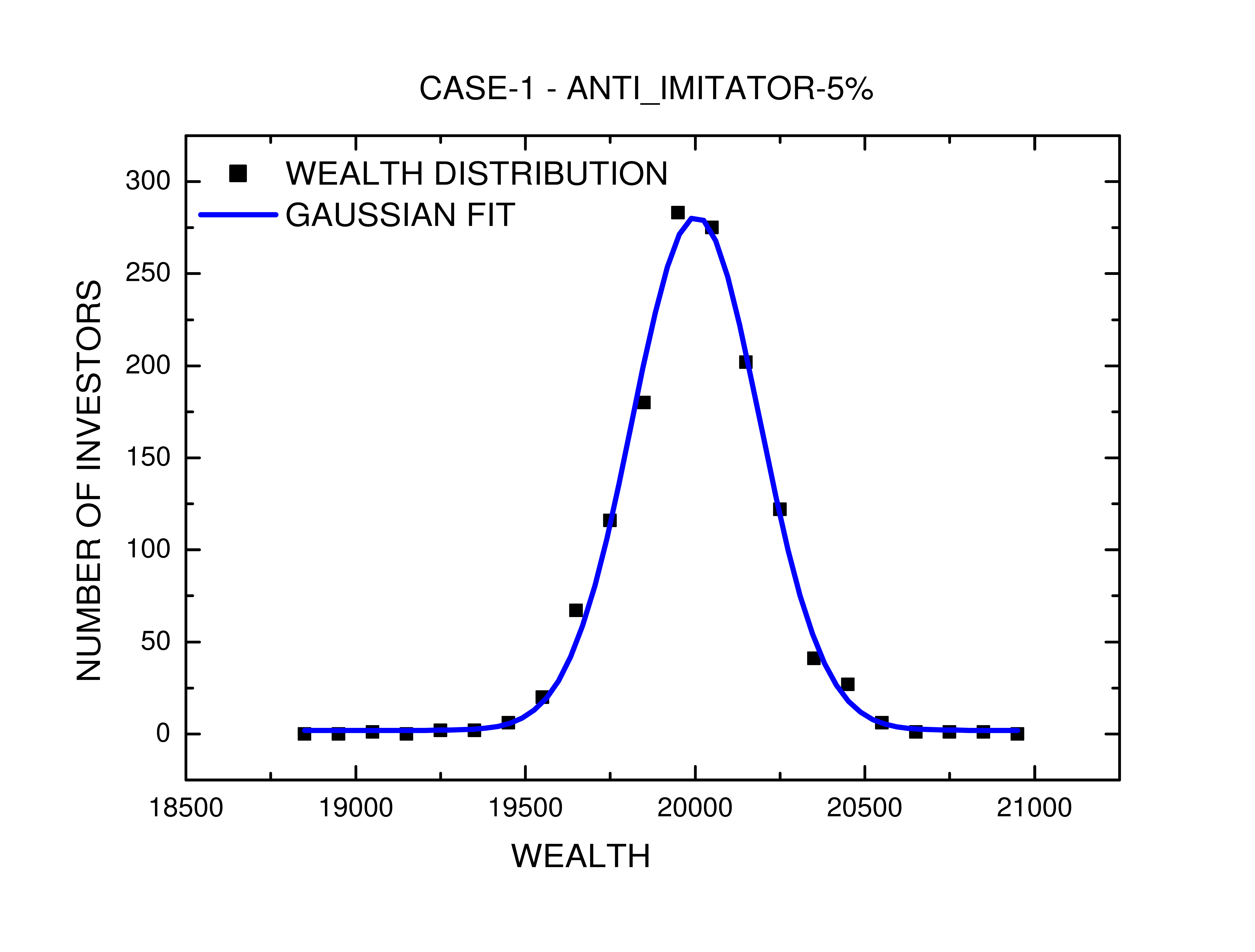}
    \includegraphics[width=0.45\linewidth]{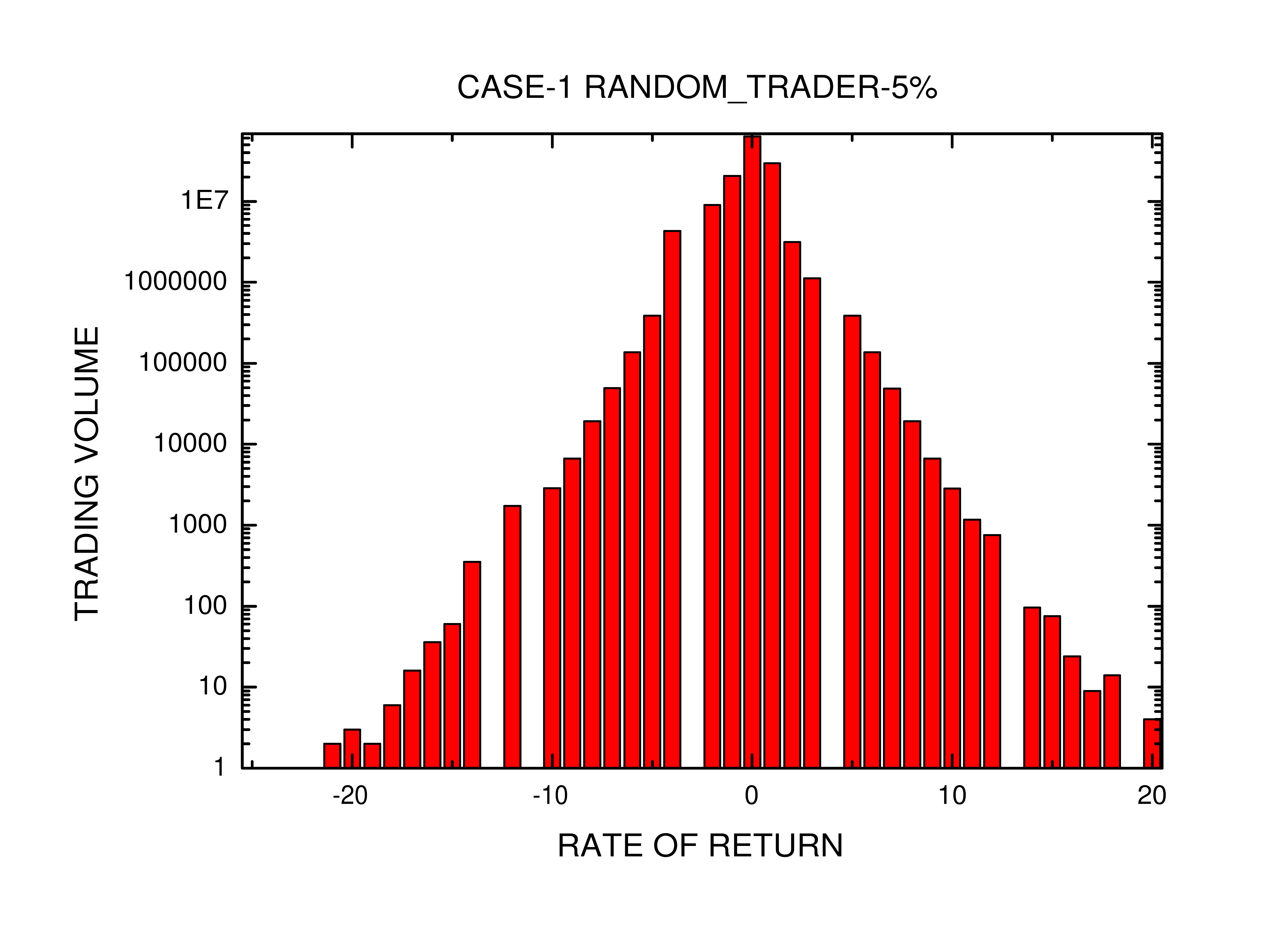}
    \includegraphics[width=0.45\linewidth]{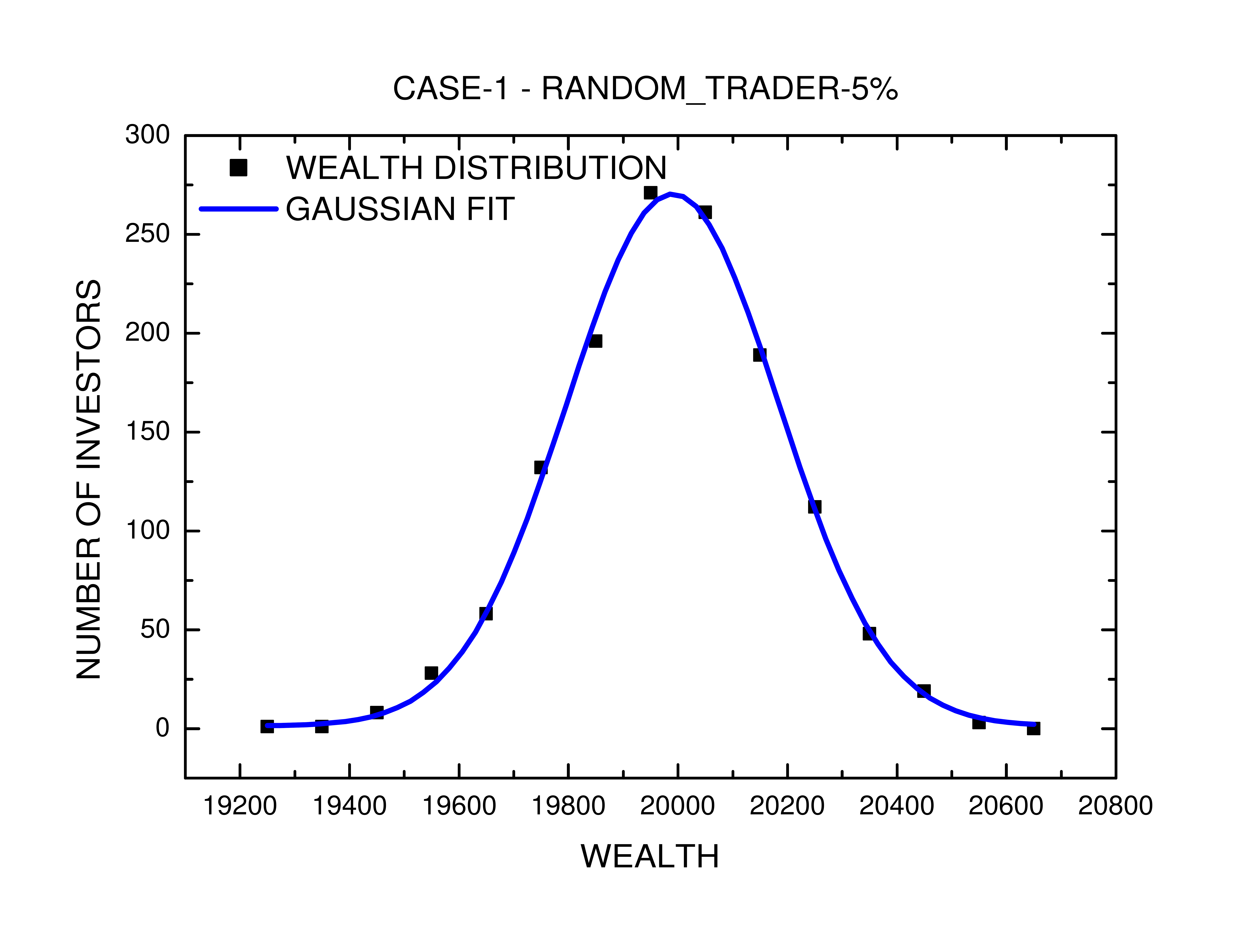}
    \includegraphics[width=0.45\linewidth]{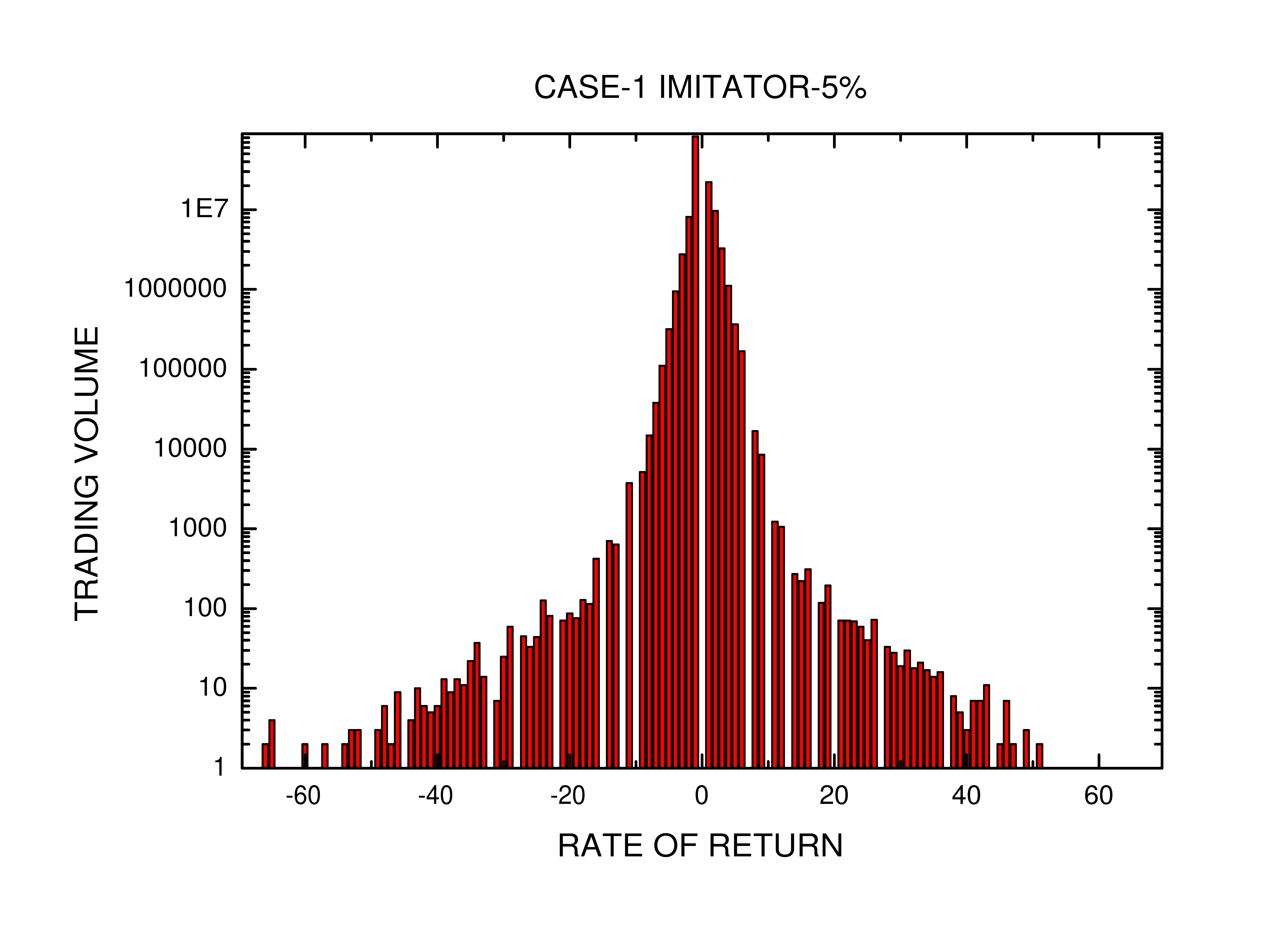}\hfill
    \includegraphics[width=0.45\linewidth]{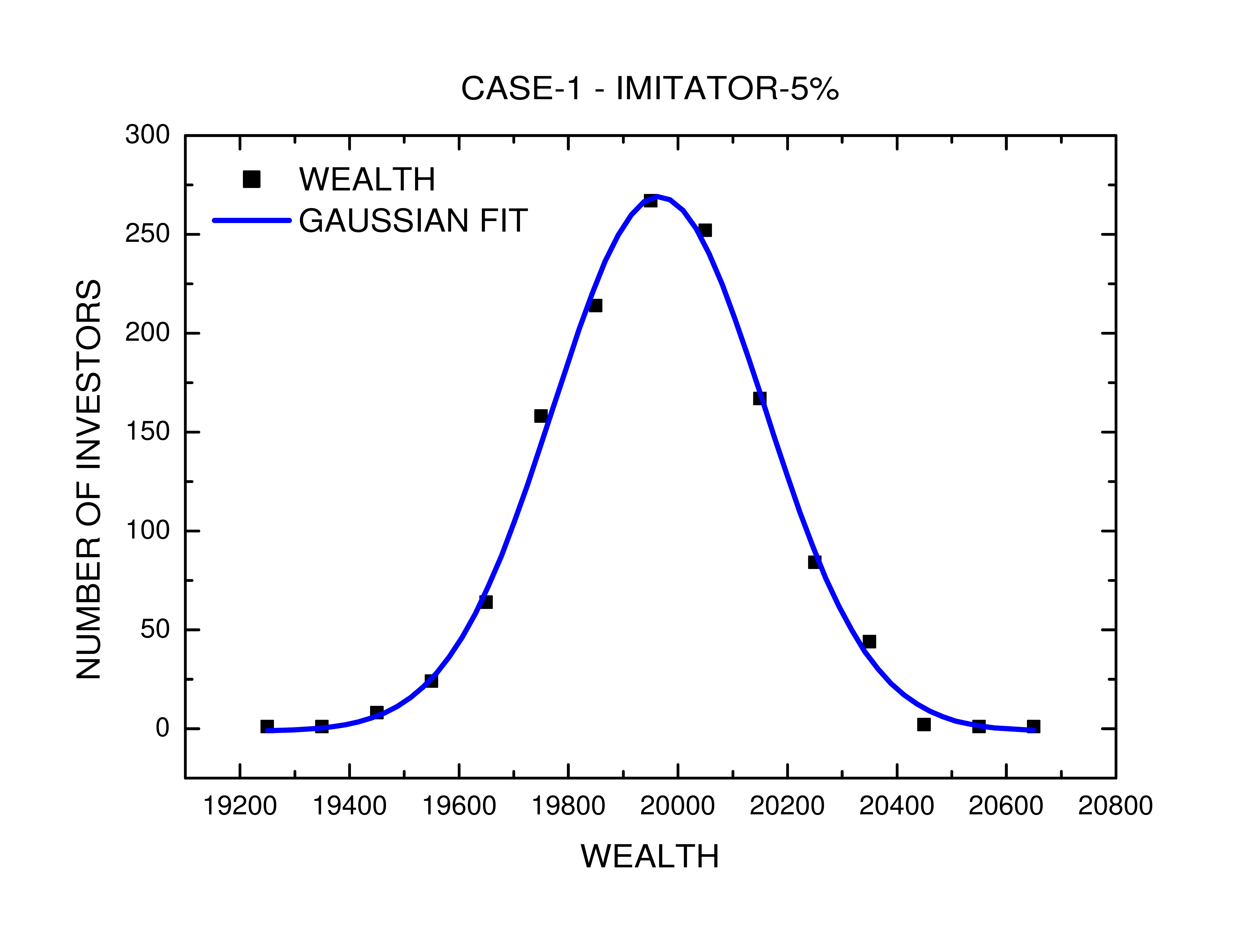}
    \caption{Rate of Return x Wealth Distribution. The figures show the results when we apply the Case-1 from the Table \ref{prob} and a probability of $5\%$ to follow the technical analysis. Left Side - Rate of Return: top-anti-imitators; middle-random-traders; bottom-imitators. Right Side - Wealth Distribution: top-anti-imitators; middle-random traders; bottom-imitators.}
    \label{return_5}
  \end{minipage}%
\end{figure*}

This section presents results about the distribution of the rate of return considering the probabilities set to follow the technical analysis (MOM). We build a probability density distribution (PDF)  computing the value of the stock for every single trading that the investor performs.  We set the algorithm-1 and algorithm-2, see Figure \ref{algor}, to cumulate all the operations over the stock market for all the investors. From the Figure \ref{return} we can see an asymmetry of the rate of return as a function of the psychological behavior. It shows us the histograms for the rate of return distribution. On the top-left we can see that the anti-imitator profile has a better return than the others profiles. The imitator profile, middle-left, has the worst one which can be seen by getting negative return. As it was to expect, on the bottom-left, we see that the random profile has a normal distribution.The remarkable result is this asymmetric distribution between the anti-imitators and imitators investors. In order to compare with the scenario where we set the probability to follow the MOM to be $5\%$, we plotted the Figure \ref{return_5}. We see that there is no impact when choosing a specific profile, because they all have a Gaussian distribution.

In order to verify the robustness of this result we considered a different algorithm for investor's decision making. Instead of comparing between two options from trust network or technical analysis, we include both strategies in a single index. For example, if the trust network of a given investors is composed of 10 agents, and 6 are holding (value +1), 3 holding (0) and 1 selling stocks (-1), the weight of the trust network for the index will be $(6+0-1)/10 = 0.5$. Supposing that technical analysis indicates a set of probabilities of $(0.8,0.1,0.1)$ to buy, sell or hold respectively, it would contribute with $0.8-0.1=0.7$ to that index, obtaining the value $1.2$. Thus, this investor will buy stocks, since the result was larger than one. Otherwise, if the result was smaller than $-1$, the investor would sell stocks. If the value was remained between $-1$ and $1$, the investor would sell ( or buy ) stocks with a probability equal to the modulus of the index, or holding otherwise. The Figure \ref{return} on the right side exhibits the results obtained with this alternative algorithm, and displays, statistically, same results as we had on the left side. However, the range of the distribution was considered enlarged, 

This asymmetry can be explained as following: when the result from the (MOM) is to buy stocks, all the imitators follow that decision which makes them spend their money at a high price. As the amount of money is limited, they hold their stocks when do not have sufficient money to buy more stocks.  On the other hand, the anti-imitators will sell their stocks at a high price. As the stock just get either more expensive or cheaper, the quantity of stock does not vanish at the same rate as the money of the imitator investors. Then, the imitators will have few stocks at a high price and the anti-imitators will have lots of money in function of the stocks which were sold. The same mechanism happens when the trend of the index is decreasing and the MOM says to sell. Thus, the imitators sell and the anti-imitators buy stocks at a price less than they have paid for. Over a period of time we have results from simulation showing that asymmetric rate of return depending on the behavior of the investors.

\begin{figure*}[htb]
   \begin{minipage}[t]{0.9\linewidth}
    \includegraphics[width=0.45\linewidth]{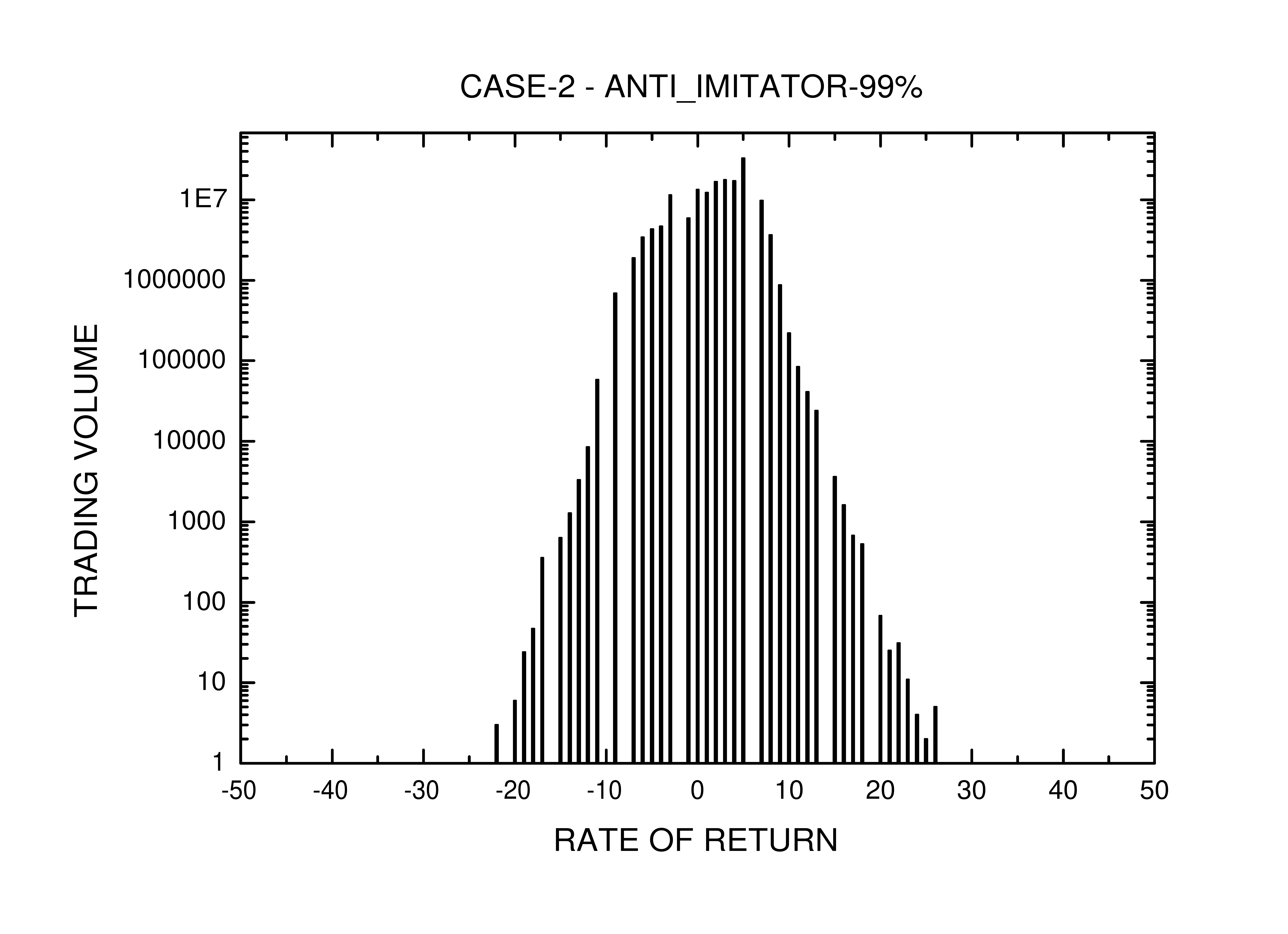}
    \includegraphics[width=0.45\linewidth]{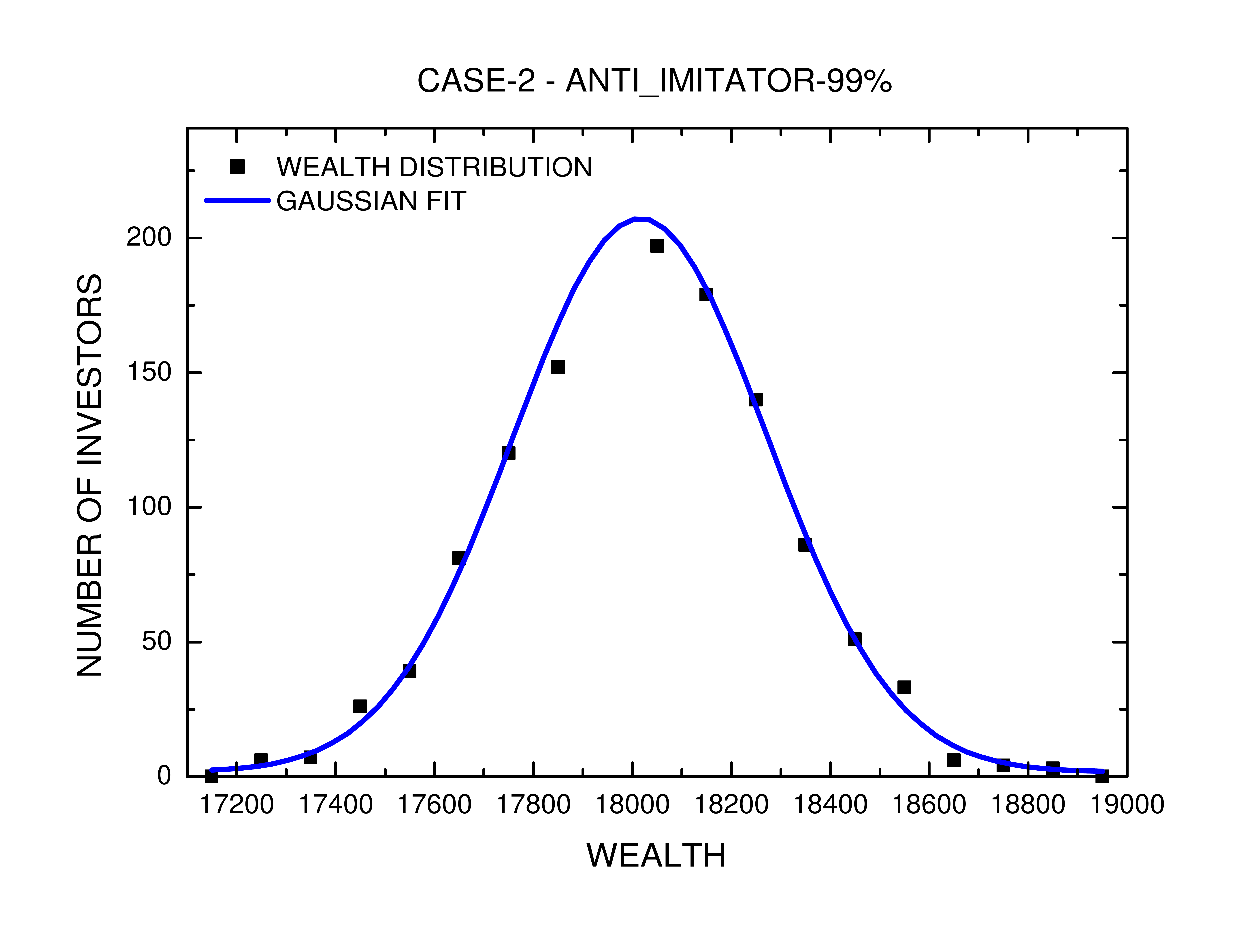}
    \includegraphics[width=0.45\linewidth]{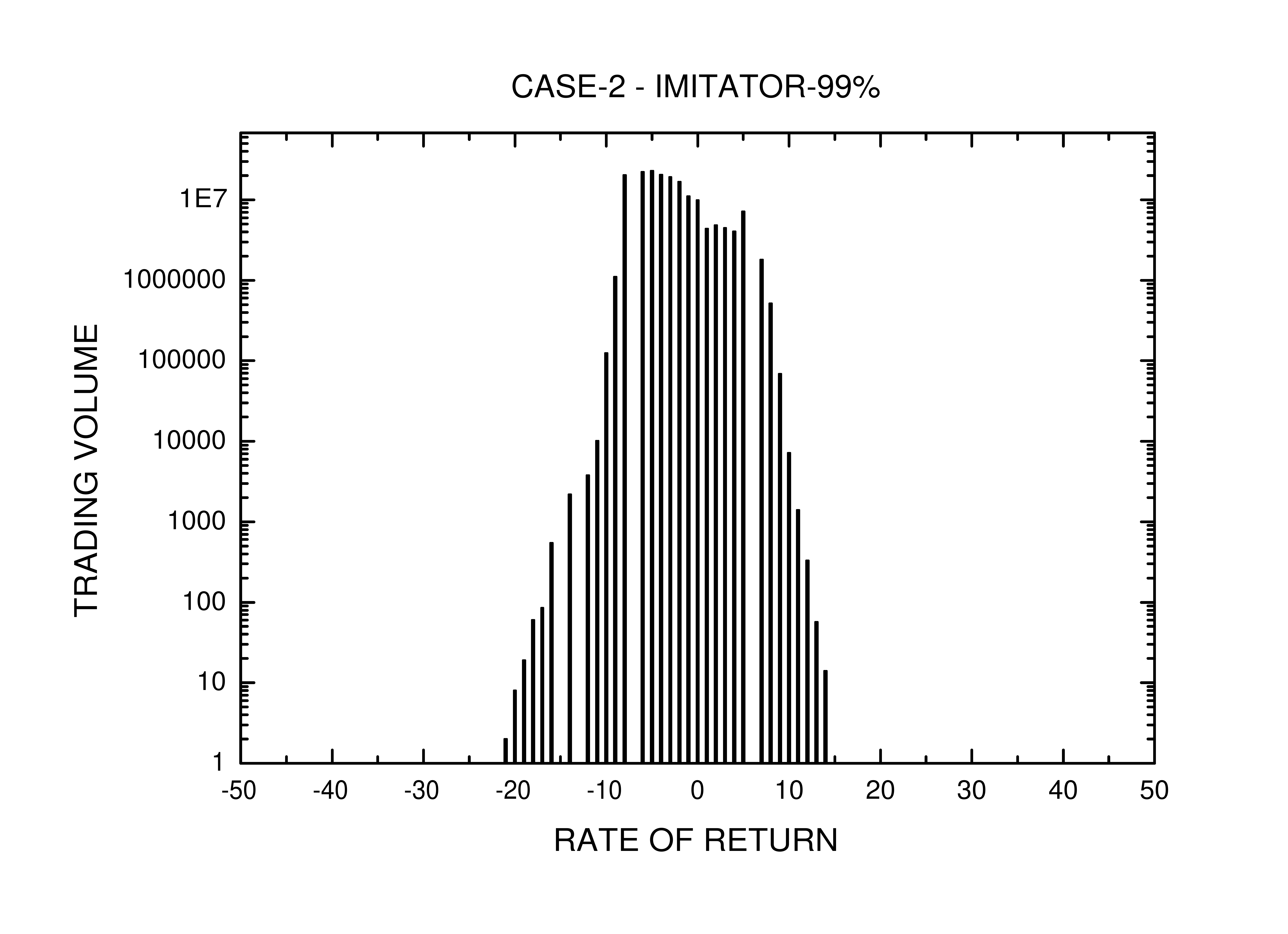}
    \includegraphics[width=0.45\linewidth]{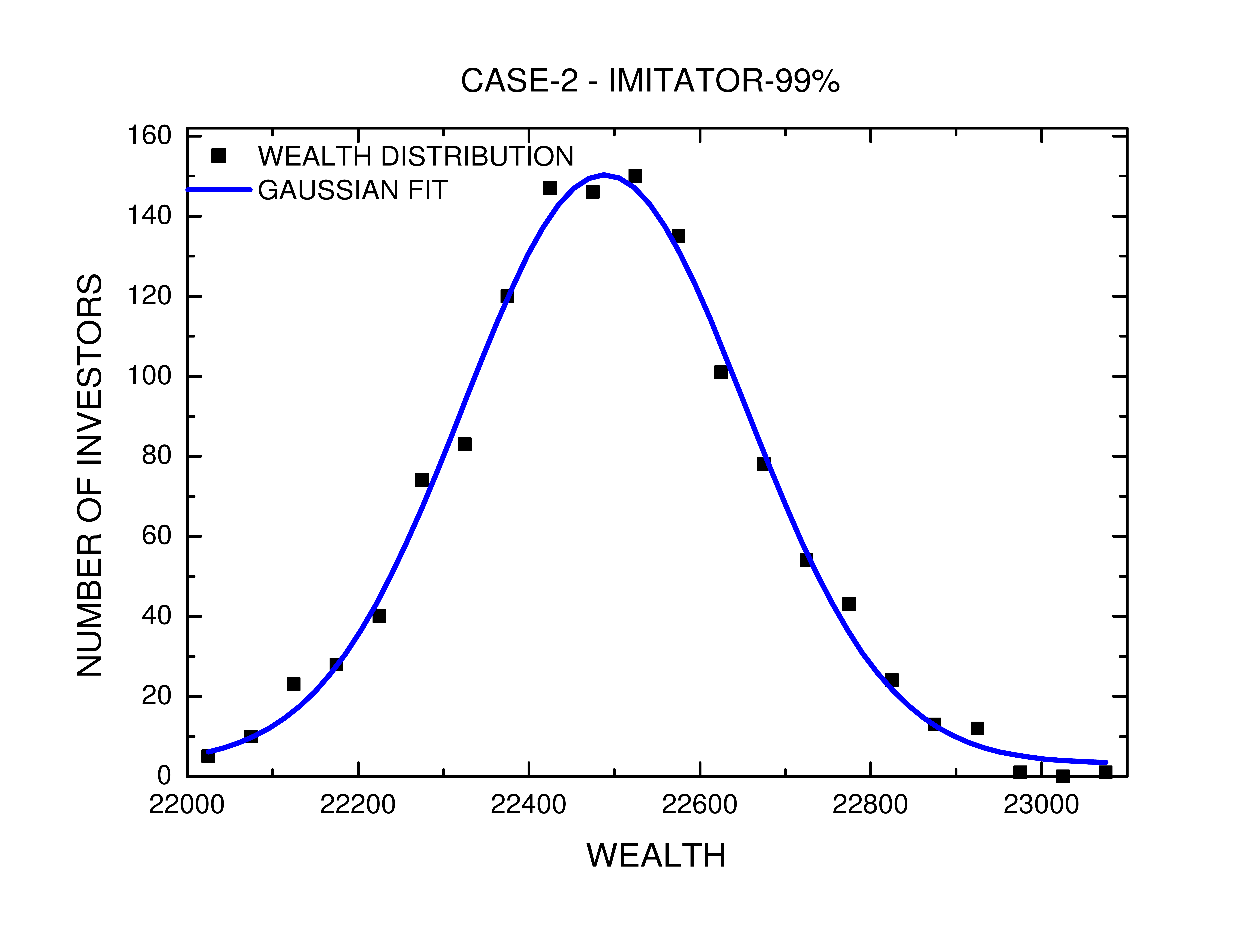}
    \includegraphics[width=0.45\linewidth]{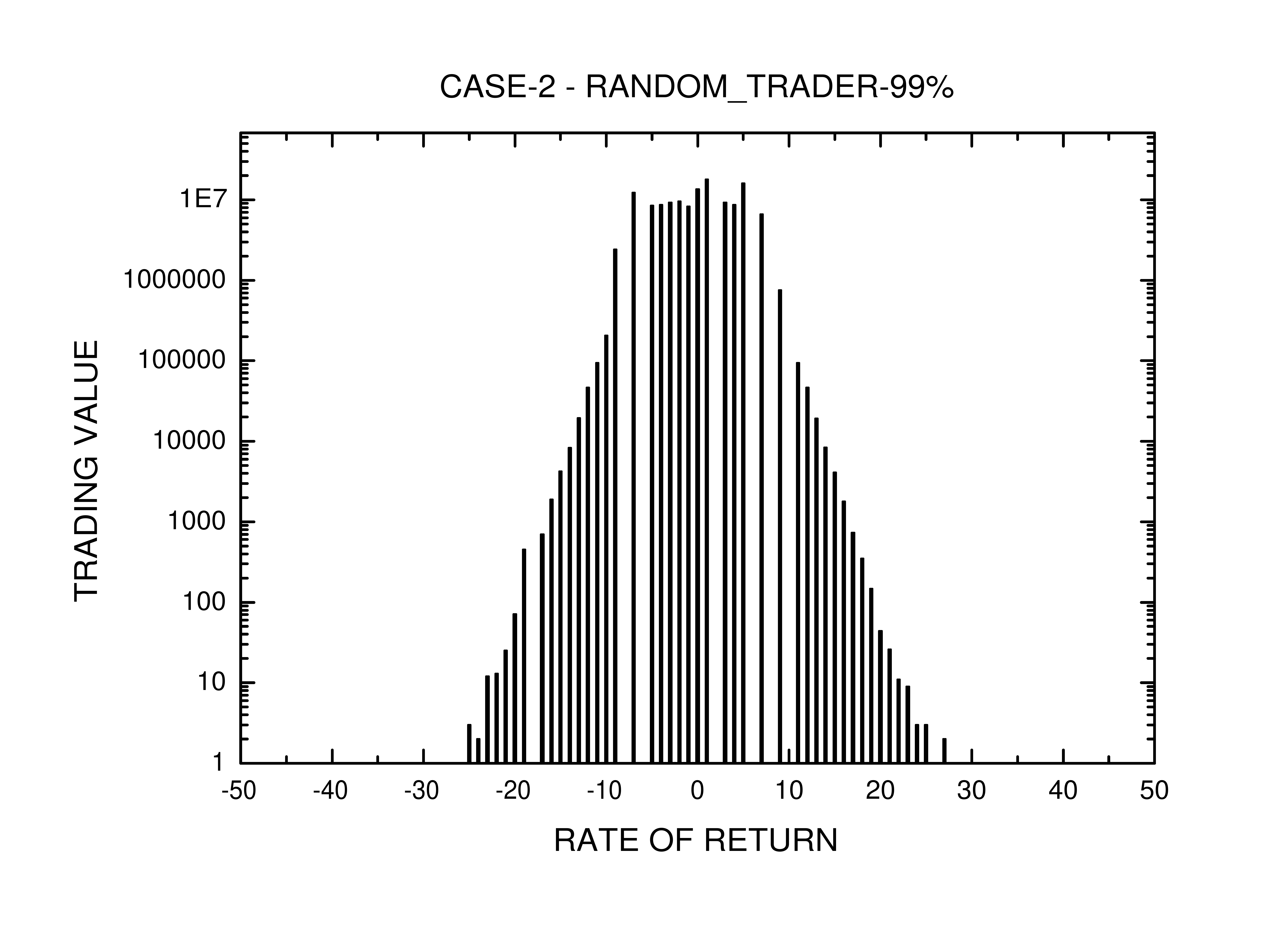}\hfill
    \includegraphics[width=0.45\linewidth]{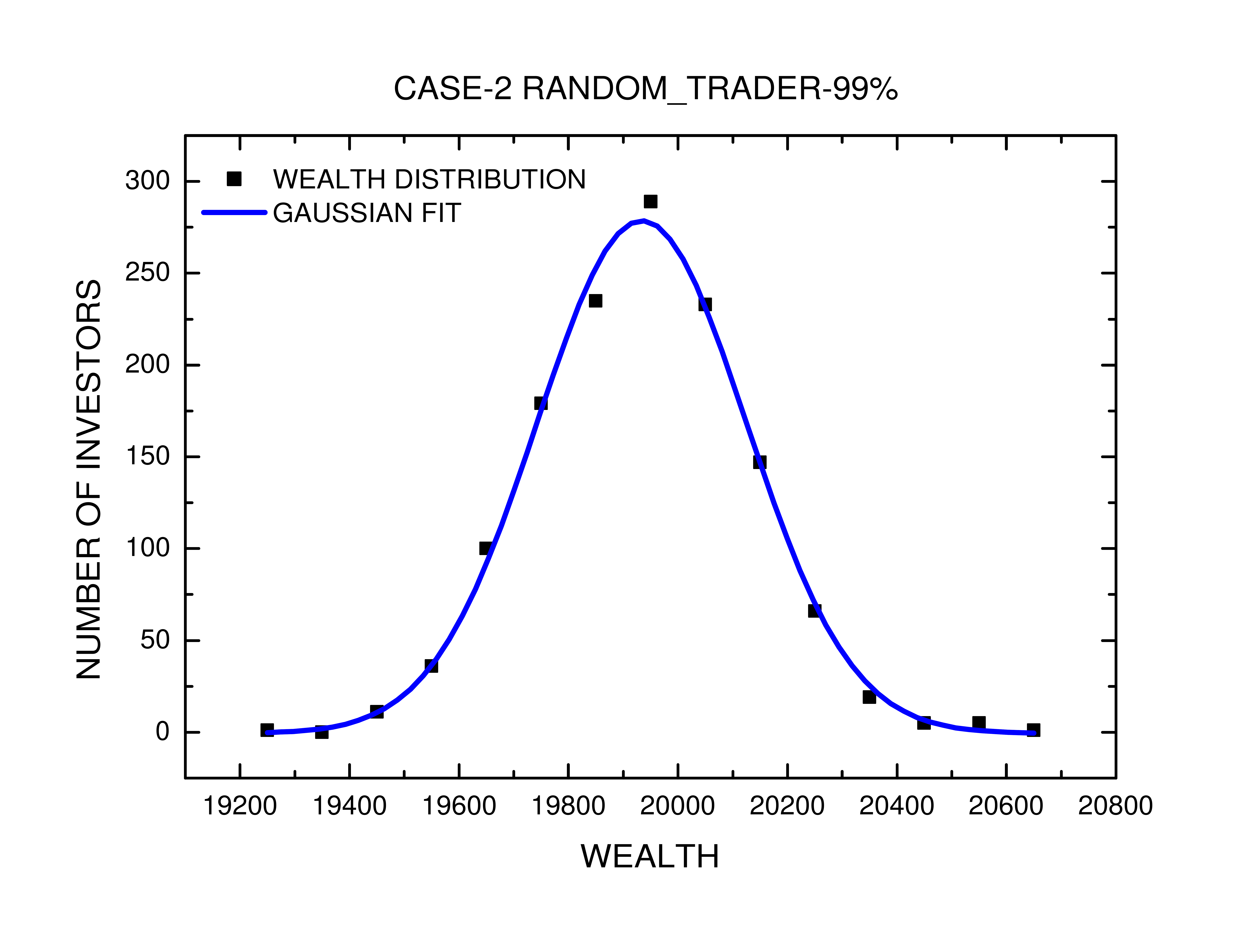}
    \caption{Rate of Return x Wealth Distribution. The figures show the results when we apply the Case-2 from the Table \ref{prob} and a probability of $99\%$ to follow the technical analysis. Left Side - Rate of Return: top-anti-imitators; middle-imitators; bottom-random traders. Right Side - Wealth Distribution: top-anti-imitators; middle-imitators; bottom-random traders.}
    \label{rateinv}
  \end{minipage}%
\end{figure*}


%
%
\begin{figure*}[htb]
  \begin{minipage}[t]{0.9\linewidth}
    \includegraphics[width=0.40\linewidth]{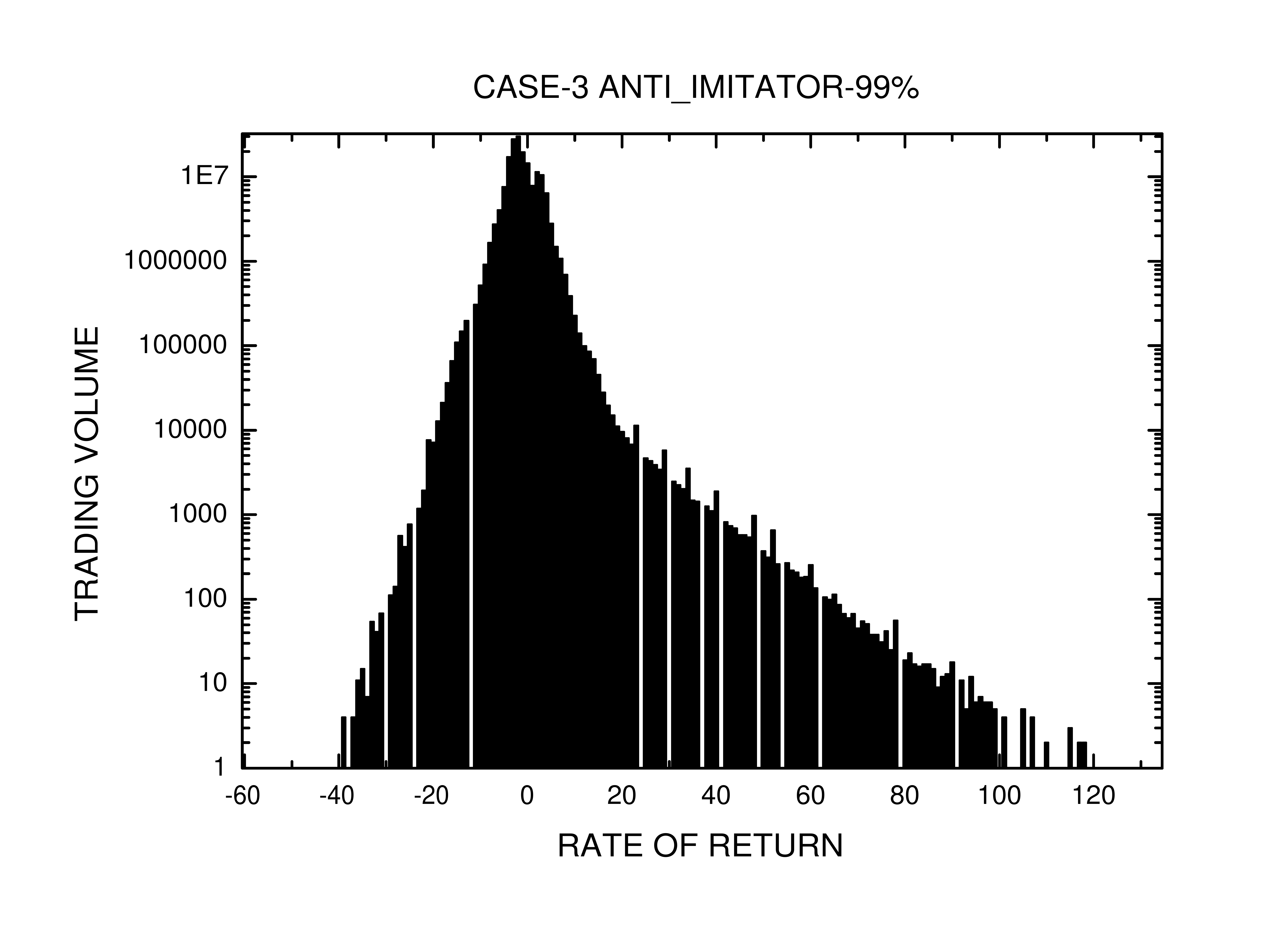}
    \includegraphics[width=0.40\linewidth]{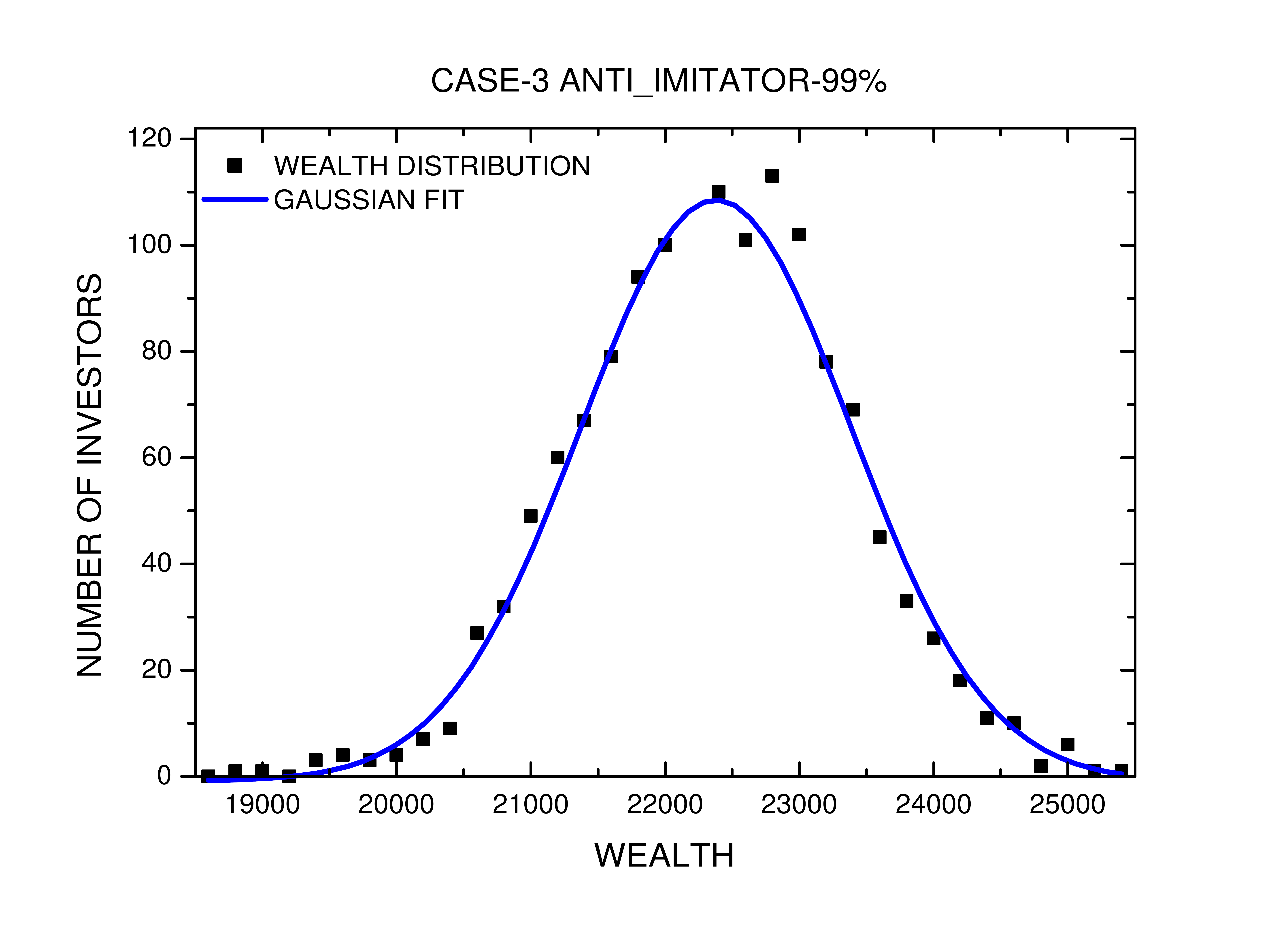}
    \includegraphics[width=0.40\linewidth]{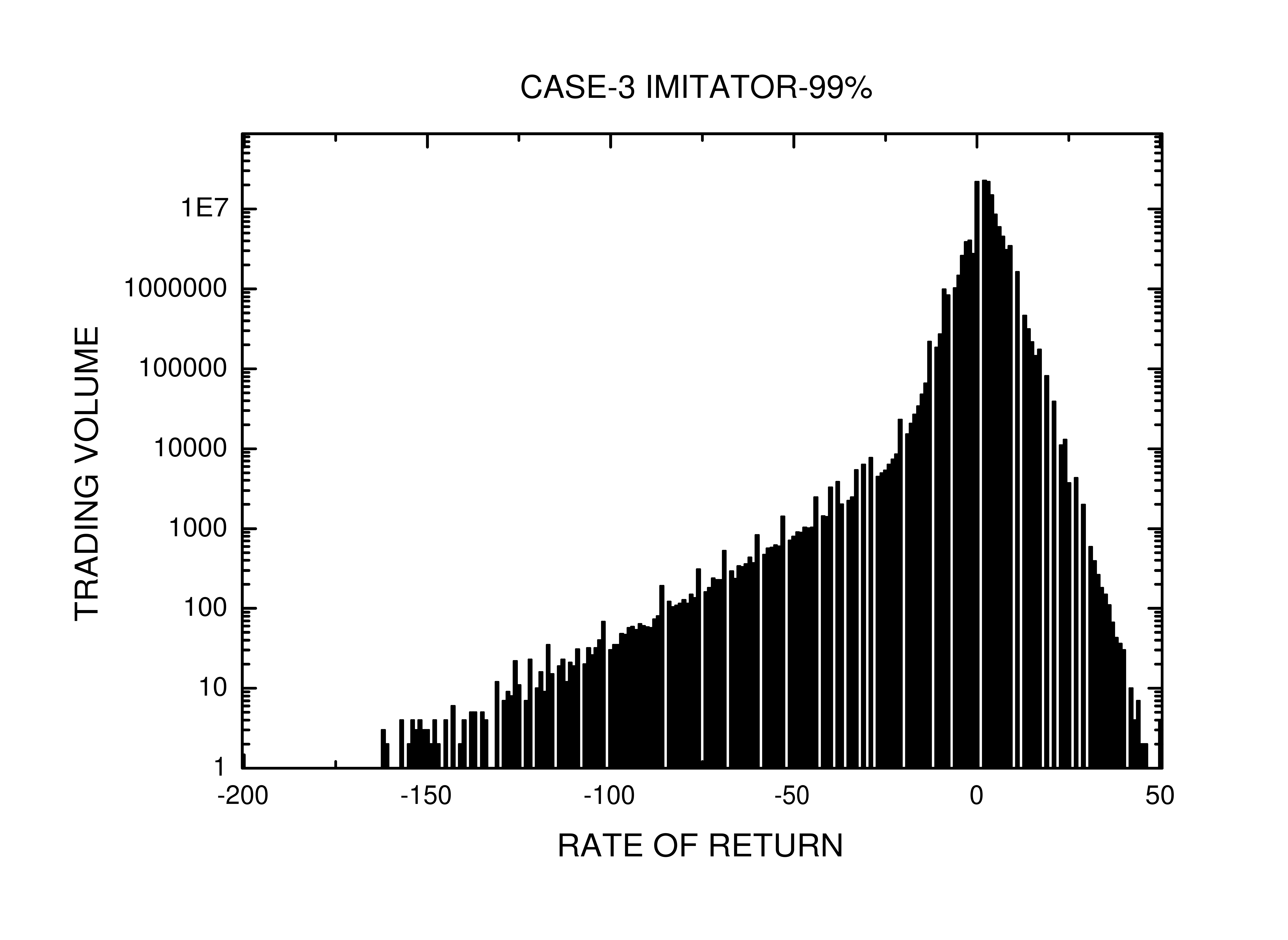}
    \includegraphics[width=0.40\linewidth]{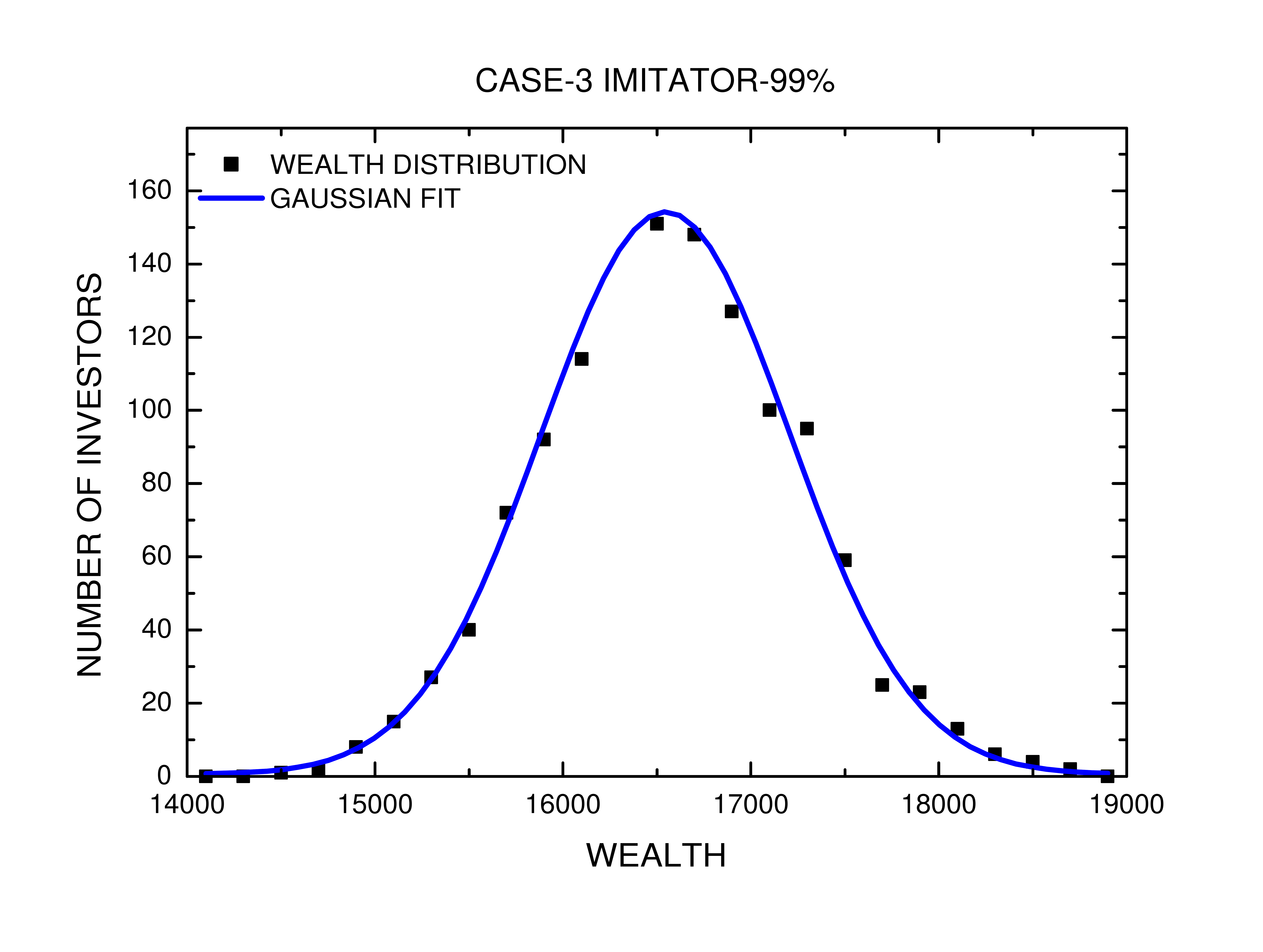}
    \includegraphics[width=0.40\linewidth]{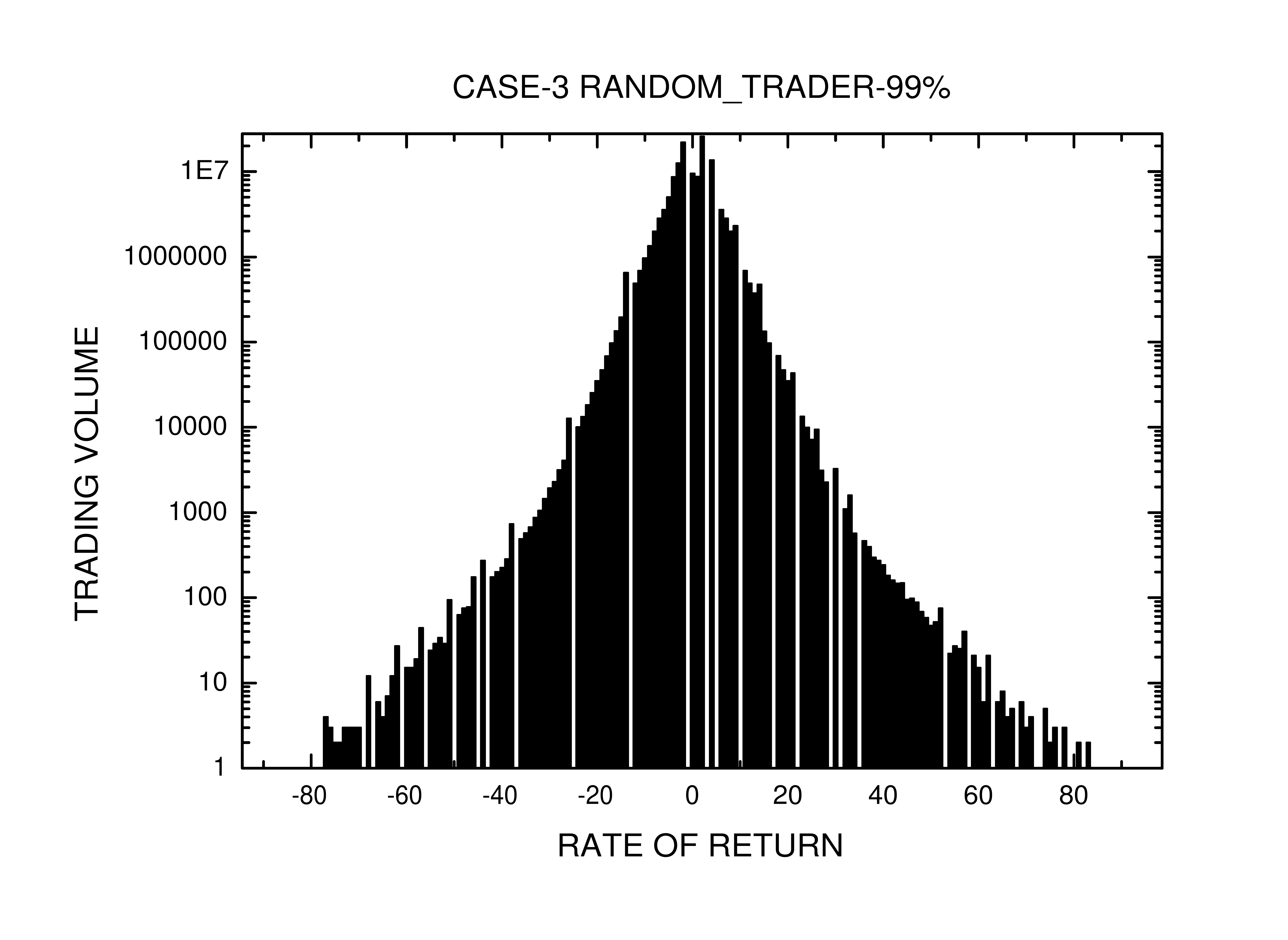}\hfill
    \includegraphics[width=0.40\linewidth]{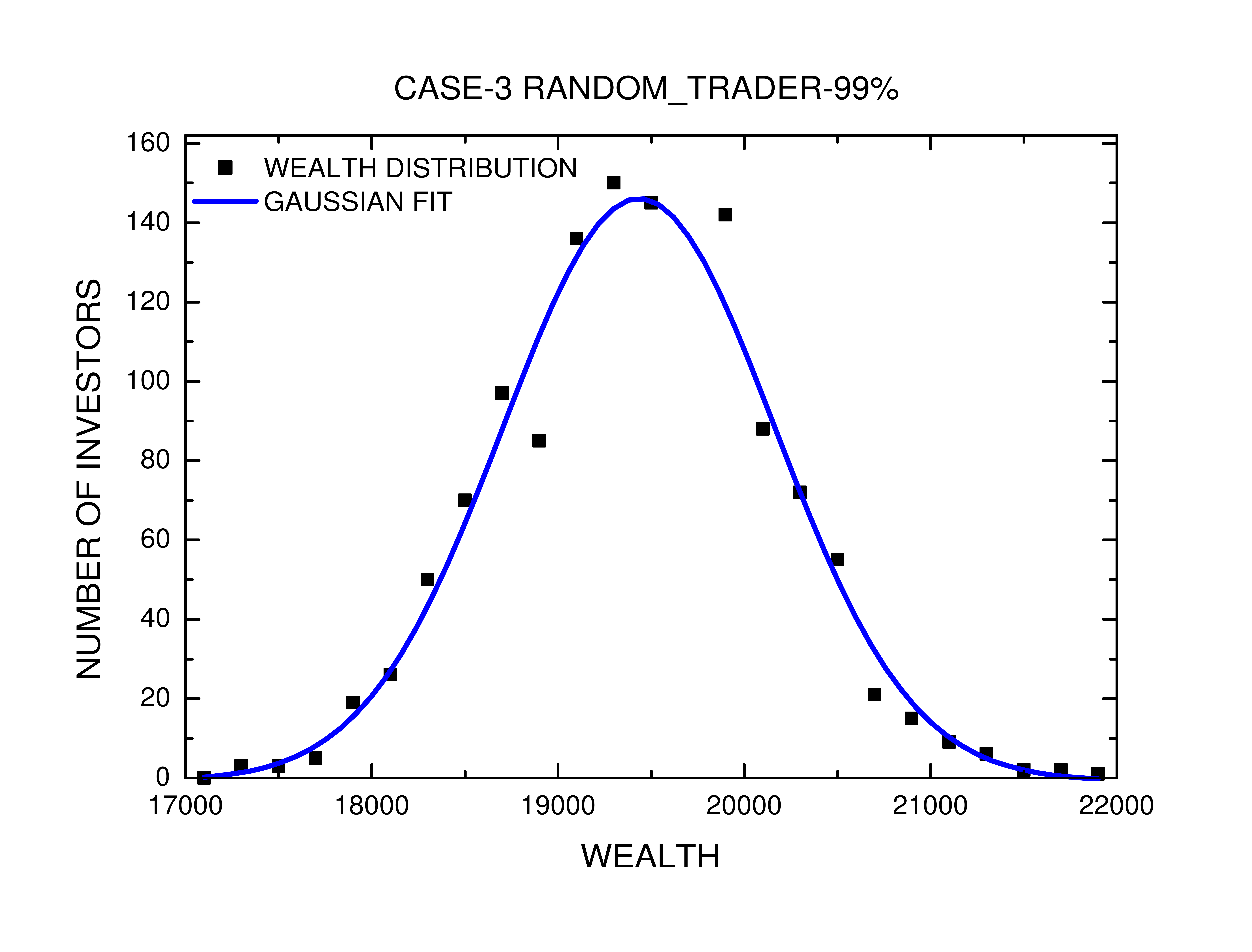}
    \caption{Rate of Return x Wealth Distribution. The figures show the results when we apply the Case-3 from the Table \ref{prob} and a probability of $99\%$ to follow the technical analysis. Left Side - Rate of Return: top-anti-imitators; middle-imitators; bottom-random traders. Right Side - Wealth Distribution: top-anti-imitators; middle-imitators; bottom-random traders.}
    \label{ratebal}
 \end{minipage}%
\end{figure*}

%
%
\section{CONCLUSION}

\begin{figure*}[htb]

    \includegraphics[width=0.40\linewidth]{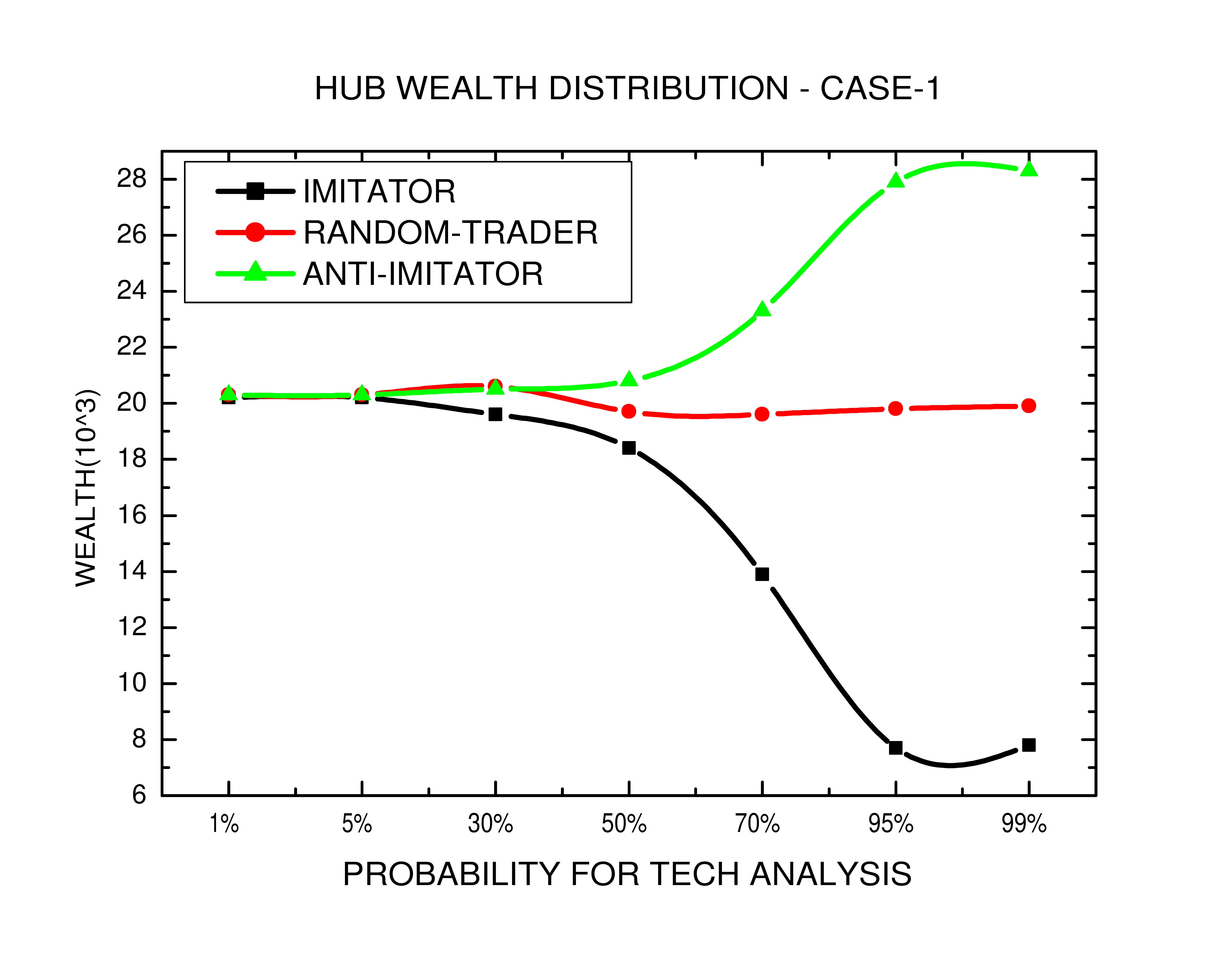}
    \includegraphics[width=0.40\linewidth]{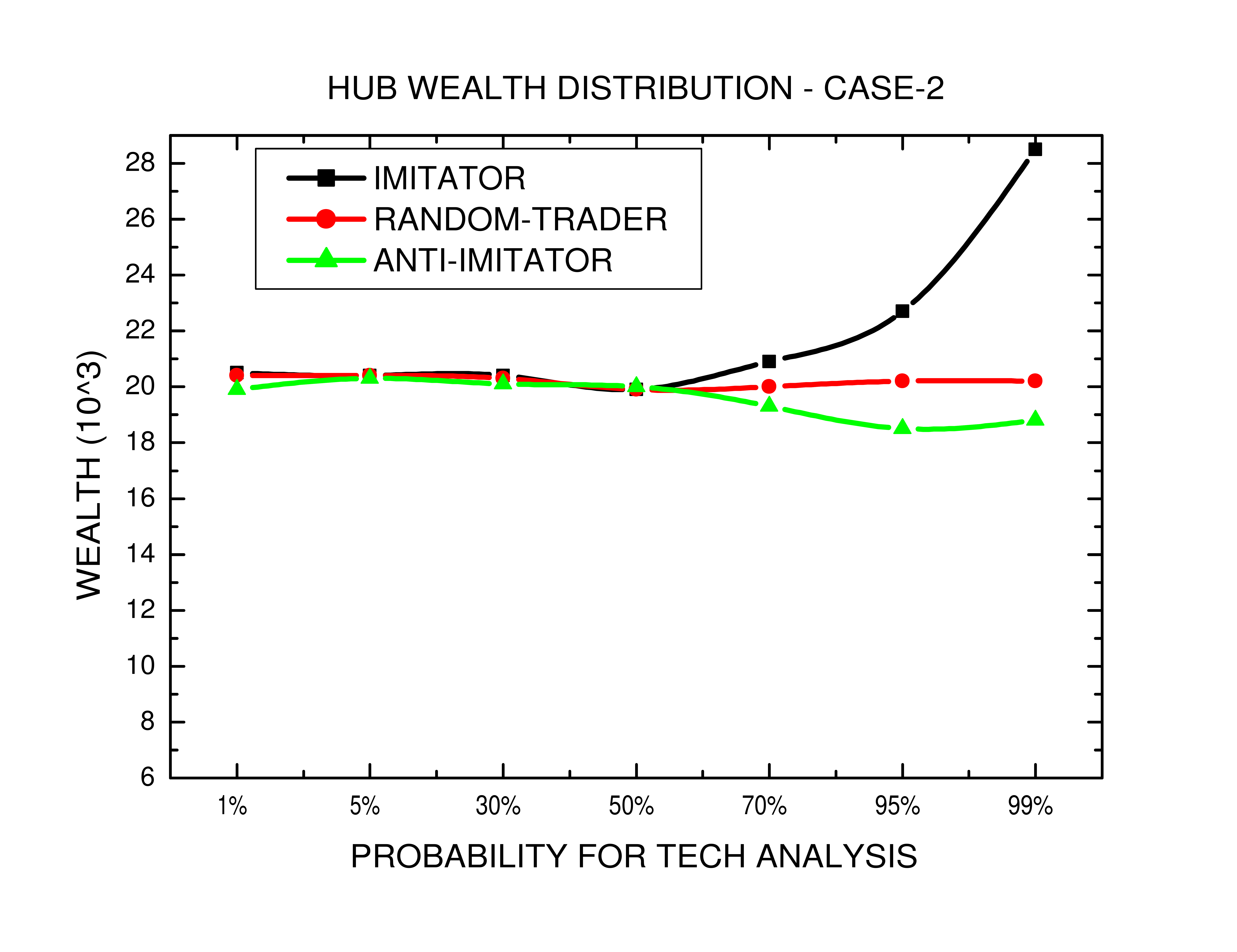}
    \includegraphics[width=0.40\linewidth]{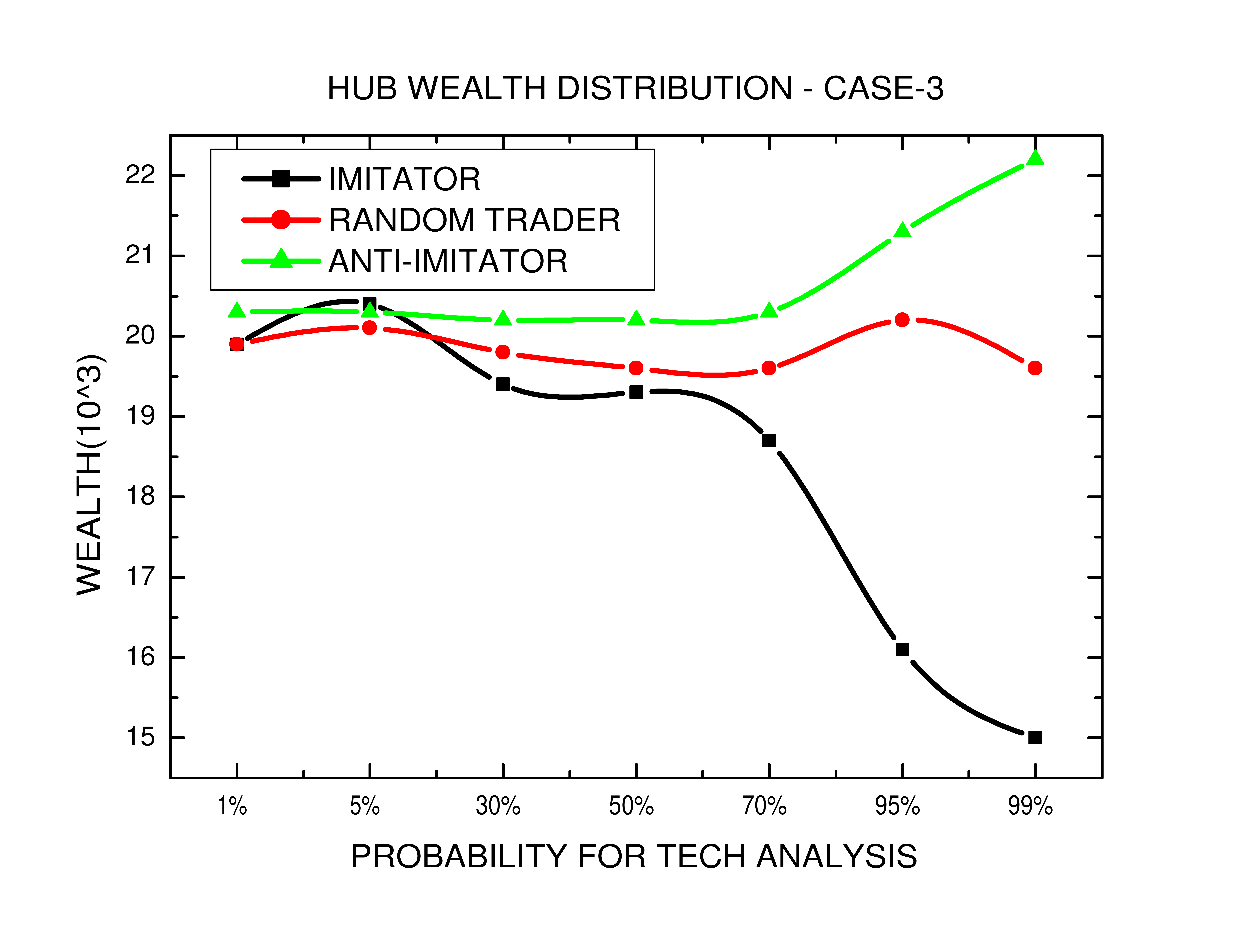}\hfill
    \includegraphics[width=0.40\linewidth]{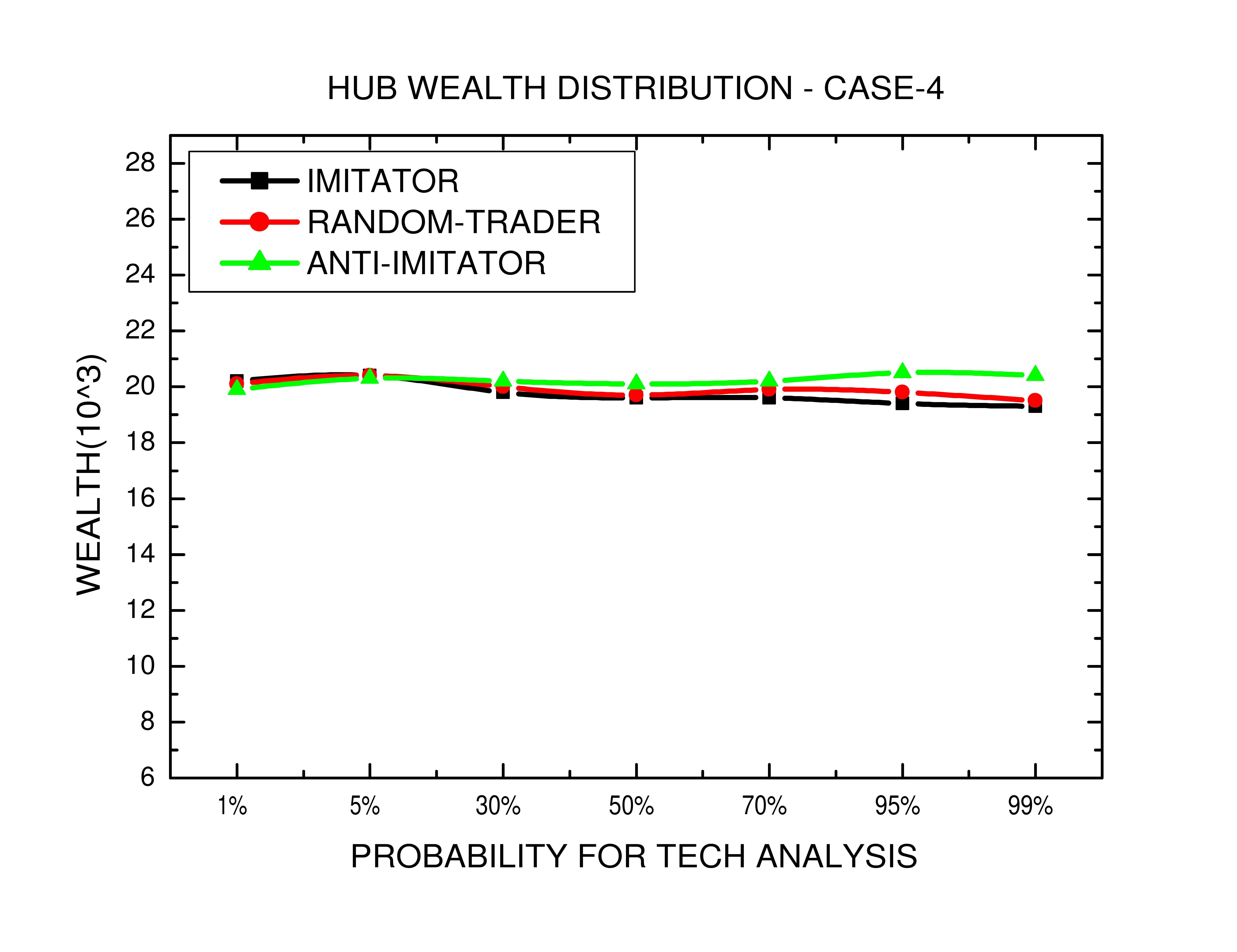}
   
   \caption{The graphics show the wealth of the Hub of the system as a function of the probability adopted to follow the technical analysis strategy for each psychological profile of the Hub. Left top: Case-1; Right top: Case-2;\newline Left bottom: Case-3; Right-bottom: Case-4 (inverted tendency of the Case-3)}
   \label{hub}
\end{figure*}

\begin{figure*}[htb]
    \centering
  \scalebox{0.70}
   {
    
    \includegraphics[width=0.80\linewidth]{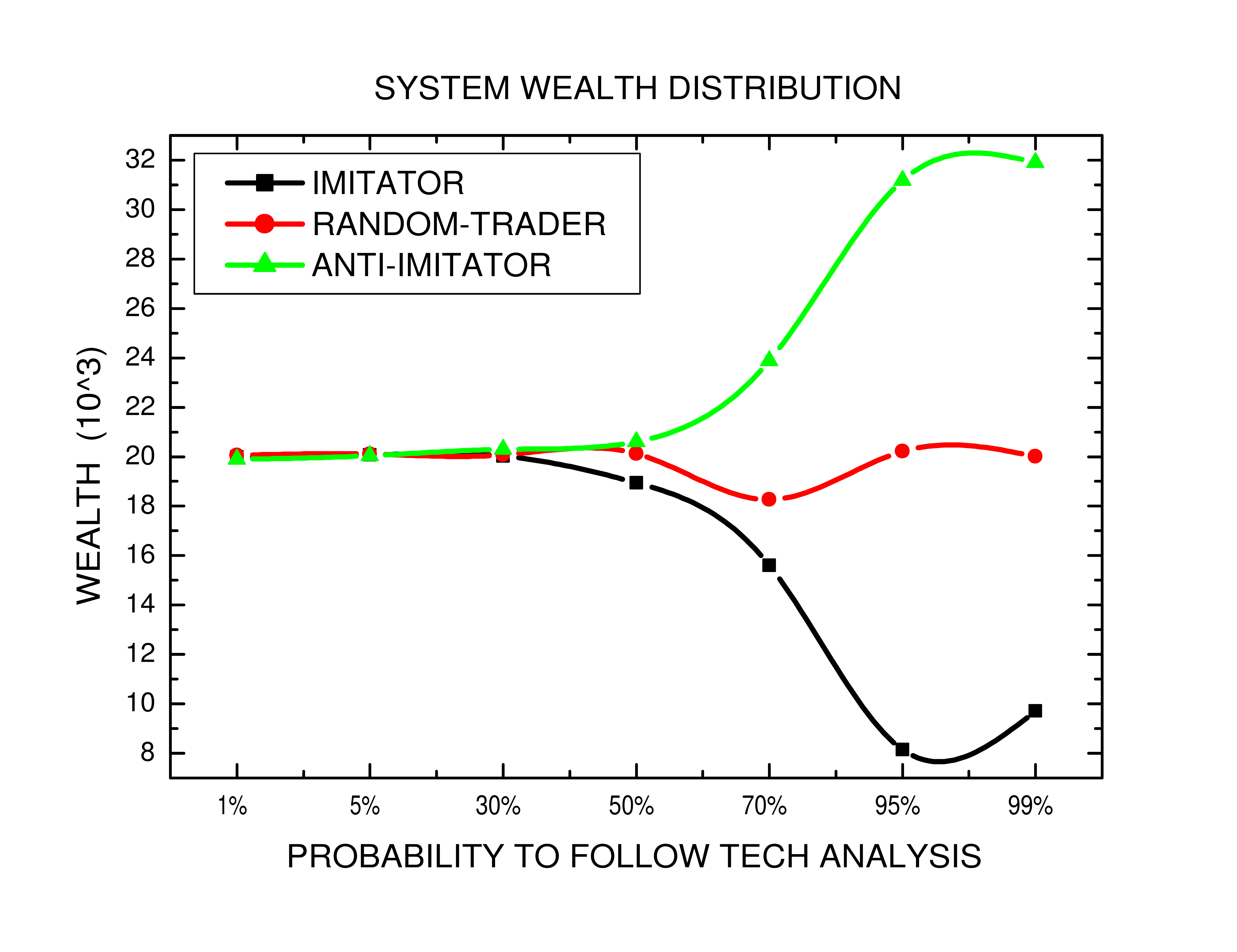}
   }
   \caption{The graphic shows shows the average wealth for every kind of psychological behavior as a function of the probability adopted to follow the technical analysis strategy applying the Case-1 from Table \ref{prob}.}
   \label{system}
\end{figure*}

\begin{figure*}[htb]
    \centering
  \scalebox{0.70}
   {
    
    \includegraphics[width=0.80\linewidth]{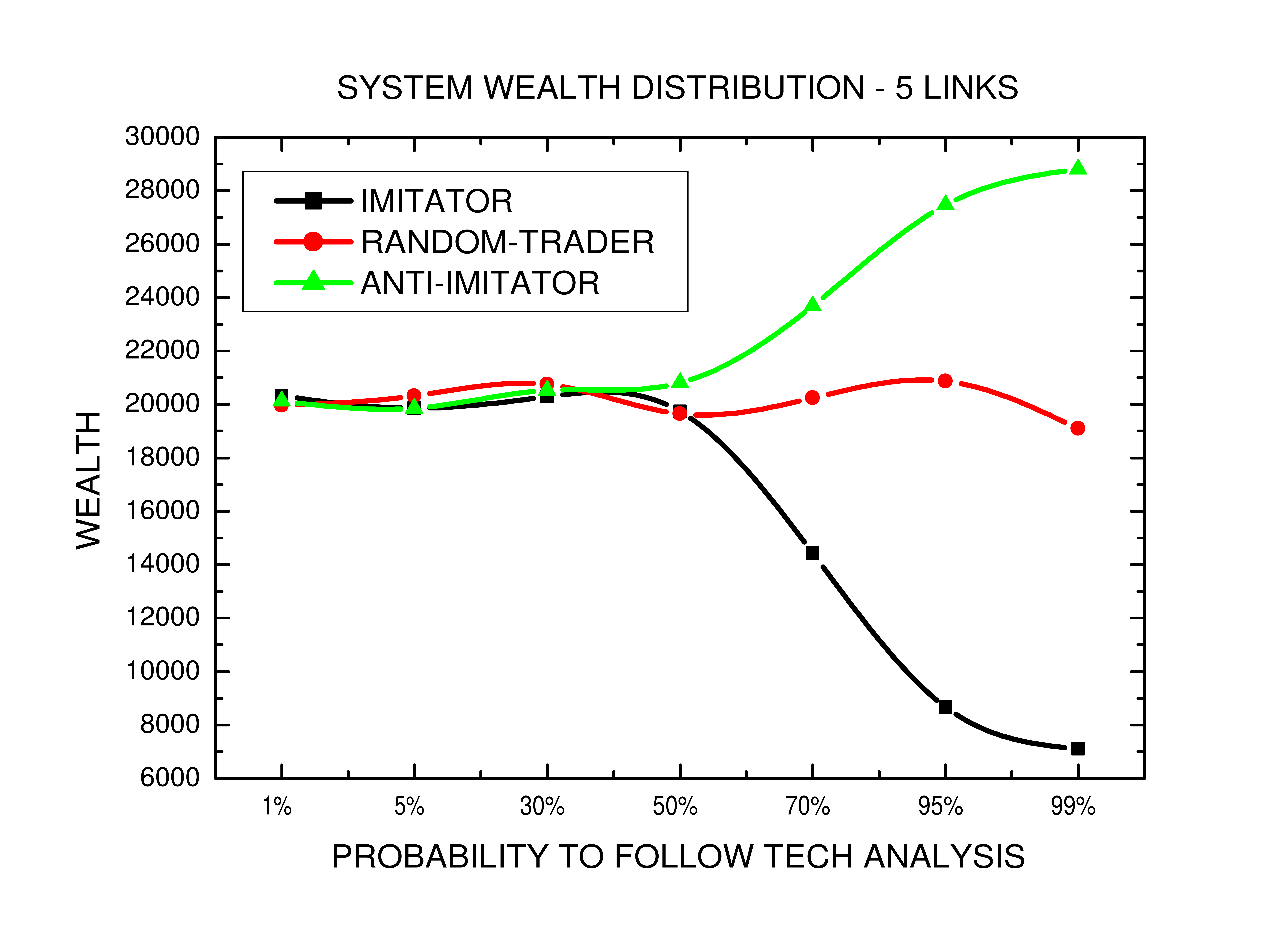}
   }
   \caption{The graphic shows the wealth of the whole system as a function of the probability adopted to follow the technical analysis strategy for each psychological profile of the investors applying the Case-1 from Table II. Each one of them shows the average value of the system when the hub was set to be anti-imitator, imitator, then random-trader.\label{link}}
   \label{system}
\end{figure*}

Analysing the result from Figure \ref{prob}, we decided to performe a new set of simulations to make this influence clearer. We create two copies of a given realization of a scale free network with one third of each behavioral profile. Then, for one copy, we chose the 300 less connected investors (5 links) to change their profiles and compare the  evolution of wealth distribution of each realization. We can observe in Figure \ref{link} that the difference between the realizations is marginal (comparing it with the Figure \ref{prob}). Then, we take again the same two copies and now we changed the behavioral profile of only one link - the hub - we compared the results for wealth distribution. It is clear that the influence of changing only  the hub is much stronger than changing those 300 less connected investors (5 links each), evincing that the hub can alter significantly the market.

In this paper we have extended a behavioral finance model in the stock market in order to study the distribution of the richness among the investors. In this sense we developed an algorithm to perform a technical analysis. By combining the two different techniques which are the neighborhood with their psychological profile as seen in our previous work \cite{Stefan} and the table which studies the oscillation of the index for different time-lags. This complex system suggests that when taking a probability greater than $50\%$ to apply the technical analysis, there is a huge chance of getting richer even though having a risk of getting the least profit which is still not too bad when comparing it with another probabilities scenarios.

The simulations results have shown us how the behavior of the investors and the technical analysis (MOM) can bring an asymmetric rate of return where the anti-imitators investors had a profitable wealth  comparing with the imitators ones. Moreover, as much as they tend to follow MOM technique as much as we can see how profitable anti-imitators investors become. From the Figures \ref{hub} and \ref{system}, we can clearly see how the anti-imitator improve his profit along the period of investment. The results from simulations, considering the random-traders investors, just confirming the results from the literature \cite{Rapisarda,Bouchaud}, which make the model robust.   

We still need a deeper study of the weight probability given for the slope between the time-lags. Furthermore, we are considering to extend the amount of the resources, a relative risk aversion and make each investor to buy and sell from each other instead of trading straight from the stock market.
\clearpage

\newpage
\bibliographystyle{elsarticle-num}
%
\appendix*
\section{Algorithm}
\label{App:algos}

\begin{figure*}[!h]
  
  \centering
   {
   \includegraphics[scale=0.8,width=0.8\linewidth]{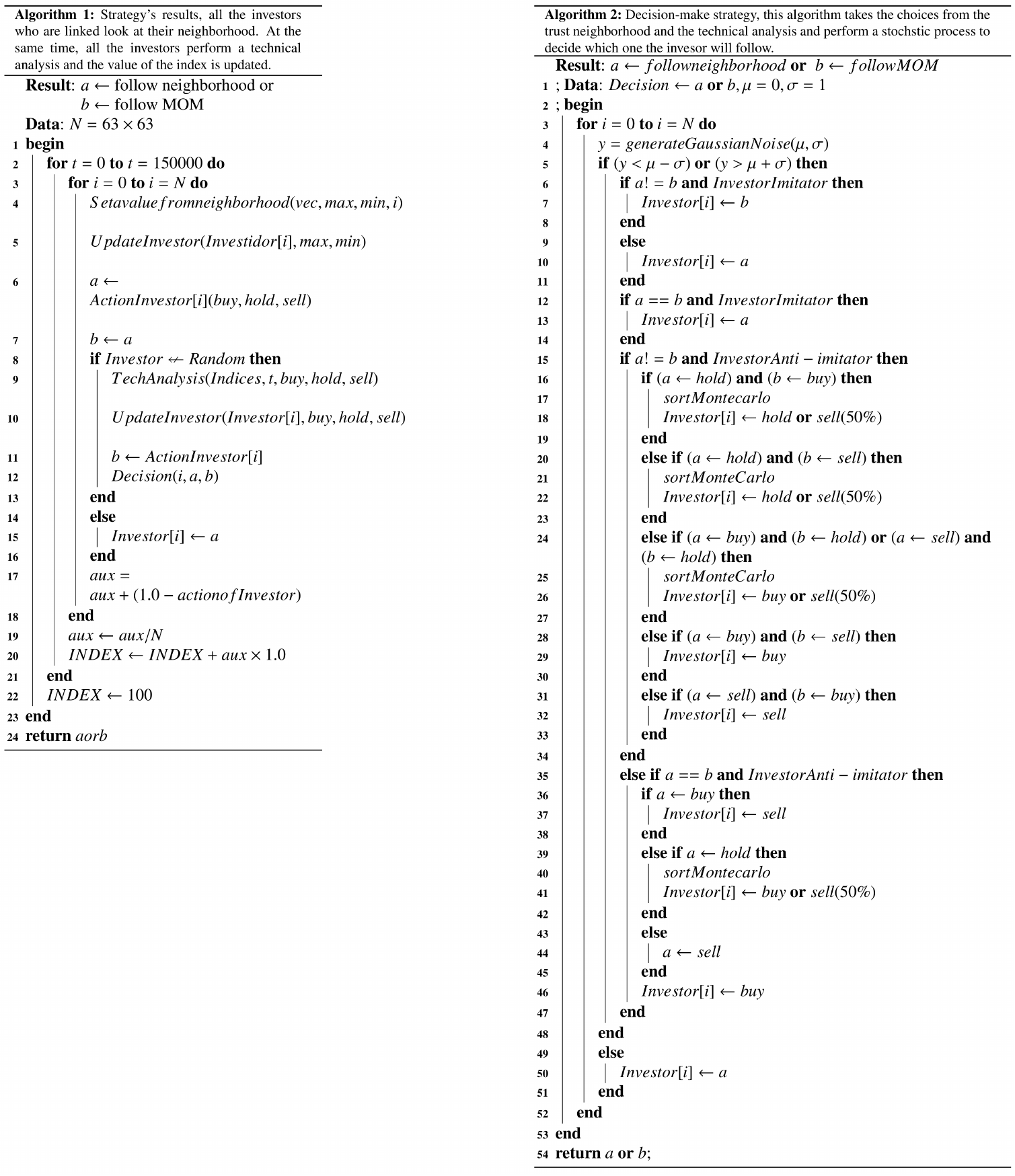}%
   }
  \caption{
  Left: The algorithm shows the strategy's results. All the investors who are linked look at their neighborhood. The investors perform a technical analysis and the value of the index is updated at the same time.
  Right: Decision-make strategy, this algorithm takes the choices from the trust neighborhood and the technical analysis and perform a stochstic process to decide which one the invesor will follow. \label{algor}}%
\end{figure*}

  The Algorithm-1, from Figure \ref{algor}, shows how we have set the process of verifying what is the state of the trust neighborhood, that means, what every single investor is performing (buying, holding, selling) at a current time and, at the same time, making each investor to perform a technical analysis over the temporal series of the index (MOM). The Algorithm-2 from \ref{algor} shows how the stochastic process works in order to decide if an investor should eithe follow the MOM result or his trust neighborhood.
  
\clearpage
\newpage
\bibliography{Qualify}

\begin{thebibliography}{10}
\expandafter\ifx\csname url\endcsname\relax
  \def\url#1{\texttt{#1}}\fi
\expandafter\ifx\csname urlprefix\endcsname\relax\def\urlprefix{URL }\fi
\expandafter\ifx\csname href\endcsname\relax
  \def\href#1#2{#2} \def\path#1{#1}\fi

\bibitem{Boccara}
N.~Boccara, Modeling Complex Systems, Springer, 2004.

\bibitem{Bouchaud}
J.~Bouchaud, M.~Potters, Theory of financial risk and derivative pricing, 2nd
  Edition, Cambridge Univ. Press, 2003.

\bibitem{lebaron06}
B.~le~Baron, IN: Handbook of Computational Economics. L. Tesfatsion and K. Judd
  eds., North-Holland, 2006, Ch. Agent-based Computational Finance, pp.
  1187--1232.

\bibitem{lebaron94}
B.~le~Baron, Chaos and nonlinear forecastability in economics and finance,
  Philosophical Transactions of the Royal Society of London (A) 348 (1994)
  397--404.

\bibitem{lebaron99}
B.~le~Baron, W.~B. Arthur, R.~Palmer, The time series properties of an
  artificial stock market, Journal of Economic Dynamics and Control 23 (1999)
  1487--1516.

\bibitem{cajueiro04}
D.~O. Cajueiro, B.~M. Tabak, The hurst exponent over time: testing the
  assertion that emerging markets are becoming more efficient., Physica. A 336
  (2004) 521--537.

\bibitem{tanyaraujo}
T.~Ara\'ujo, F.~Lou\c{c}\~{a}, Modeling a multi-agents system as a network,
  International Journal of Agent Technologies and Systems 1 (2009) 17--29.

\bibitem{fudenberg}
D.~Fudenberg, Game theory, MIT Press, Cambridge, 1991.

\bibitem{Hart}
M.~Hart, P.~Jefferies, N.~F. Johnson, P.~M. Hui, Crowd–anticrowd theory of
  the minority game, Physica A: Statistical Mechanics and its Applications
  298~(3) (2001) 537--544.

\bibitem{Coolen}
A.~C.~C. Coolen, N.~Shayeghi, Generating functional analysis of minority games
  with inner product strategy definitions, Journal of Physics A: Mathematical
  and Theoretical 41~(32) (2008) 324005.

\bibitem{Lux2005}
T.~Lux, Emergent Statistical Wealth Distributions in Simple Monetary Exchange
  Models: A Critical Review, Springer Milan, 2005, pp. 51--60.

\bibitem{Lux2009}
T.~Lux, {Applications of statistical physics in finance and economics}, 2009.

\bibitem{Lux2012}
T.~Lux, Estimation of an agent-based model of investor sentiment formation in
  financial markets, Journal of Economic Dynamics and Control 36~(8) (2012)
  1284 -- 1302.

\bibitem{Mitchell}
M.~Mitchell, Complexity: A Guided Tour, Oxford University Press, Inc., New
  York, NY, USA, 2009.

\bibitem{sornette}
D.~Sornette, Why Stock Markets Crash - Critical Events in Complex Financial
  Systems, Princeton University Press, 2003.

\bibitem{mantegna}
R.~N. Mantegna, H.~E. Stanley, {Introduction to Econophysics: Correlations and
  Complexity in Finance}, 0th Edition, Cambridge University Press, 1999.

\bibitem{kirman10}
A.~Kirman, Complex Economics: Individual and Collective Rationality, Routledge,
  2010.

\bibitem{Amos}
A.~Tversky, D.~Kahneman, The framing of decisions and the psychology of choice,
  Science 211 (1981) 453--458.

\bibitem{Fabozzi}
F.~F. J., M.~F. P., J.~Frank, Foundations of Financial Markets and
  Institutions, 4th Edition, Pearson Education (US), 2009.

\bibitem{Lux1}
T.~Lux, The socio-economic dynamics of speculative markets: interacting agents,
  chaos, and the fat tails of return distributions, Journal of Economic
  Behavior \& Organization 33~(2) (1998) 143 -- 165.

\bibitem{Strogatz}
M.~E.~J. Newman, S.~H. Strogatz, D.~J. Watts, Random graphs with arbitrary
  degree distribution and their applications, Physical Review E 64 (2001)
  026118.

\bibitem{Newman}
M.~E.~J. Newman, Models of the small world, Journal of Statistical Physics
  101~(3/4) (2000) 819--841.

\bibitem{cohen2010}
R.~Cohen, S.~Havlin, Complex networks: structure, robustness and function,
  Cambridge university press, 2010.

\bibitem{Philip}
P.~Ball, Why society is a complex matter meeting twenty-first century
  challenges with a new kind of science, Springer, Berlin New York, 2012.

\bibitem{santos}
F.~C. Santos, J.~M. Pacheco, Scale-free networks provide a unifying framework
  for the emergence of cooperation, Phys. Rev. Lett. 95 (2005) 098104.

\bibitem{Barabasi}
R.~Albert, A.-L. Barab{\`a}si, Statistical mechanics of complex networks,
  Reviews of Modern Physics 74 (2002) 47--97.

\bibitem{Tiziana1}
M.~Tumminello, T.~Aste, T.~{Di Matteo}, R.~N. Mantegna, {A tool for filtering
  information in complex systems}, Proceedings of the National Academy of
  Sciences USA 102 (2005) 10421--10426.

\bibitem{Bonan}
B.~Hou, Y.~Yao, D.~Liao, Identifying all-around nodes for spreading dynamics in
  complex networks, Physica A: Statistical Mechanics and its Applications
  391~(15) (2012) 4012 -- 4017.

\bibitem{Newman_2000}
M.~E.~J. Newman, Models of the small world, Journal of Statistical Physics
  101~(3) (2000) 819--841.
\newblock \href {http://dx.doi.org/10.1023/A:1026485807148}
  {\path{doi:10.1023/A:1026485807148}}.

\bibitem{Stefan}
F.~M. Stefan, A.~P.~F. Atman, Is there any connection between the network
  morphology and the fluctuations of the stock market index?, PHYSICA
  A-Statistical Mechanics and its Applications 419 (2015) 630--641.

\bibitem{Steven}
S.~Achelis, Technical Analysis from A to Z, 4th Edition, McGraw-Hill, 2000.

\bibitem{Pring}
M.~Pring, Technical Analysis Explained: The Successful Investor's Guide to
  Spotting Investment Trends and Turning Points, McGraw-Hill Education, 2002.

\bibitem{Murphy}
J.~J. Murphy, J.~J. Murphy, Technical analysis of the financial markets, New
  York Institute of Finance, Fishkill, N.Y., 1999.

\bibitem{Edwards}
R.~D. Edwards, J.~Magee, Technical analysis of stock trends / by Robert D.
  Edwards and John Magee, [5th ed.]. Edition, J. Magee Springfield, Mass, 1969.

\bibitem{Irwin}
S.~H. Irwin, C.-H. Park, What do we know about the profitability of technical
  analysis?, Social Science Research Network Working Paper Series.

\bibitem{Rapisarda}
A.~E. Biondo, A.~Pluchino, A.~Rapisarda, D.~Helbing, {A}re {R}andom {T}rading
  {S}trategies {M}ore {S}uccessful than {T}echnical {O}nes?, PLoS one 8~(7)
  (2013) e68344--.

\bibitem{barabasi99}
A.-L. Barab{\`a}si, R.~Albert, Emergence of scaling in random networks, Science
  286 (1999) 509--512.

\bibitem{lebaron00}
B.~LeBaron, {Agent-based computational finance: Suggested readings and early
  research}, Journal of Economic Dynamics and Control 24~(5-7) (2000) 679--702.

\end{thebibliography}
\end{document}